\documentclass[journal]{IEEEtran}
\usepackage{xr-hyper}
\usepackage[table,xcdraw]{xcolor}
\usepackage[
    colorlinks=true,
    linkcolor=black,
    anchorcolor=black,
    citecolor=black,
    filecolor=black,
    menucolor=black,
    runcolor=black,
    urlcolor=black
]{hyperref}
\usepackage{algorithm}
\usepackage{algpseudocode}
\usepackage{amsmath,amssymb}
\usepackage[T1]{fontenc}
\usepackage{microtype,array,placeins}
\usepackage{xurl}
\usepackage{mathtools}
\usepackage{cleveref}
\usepackage{lineno}
\usepackage{orcidlink}
\usepackage{caption} %
\usepackage{animate}
\usepackage{extarrows}
\usepackage{tabularx}
\usepackage{booktabs}
\usepackage[utf8]{inputenc}
\usepackage{textalpha}
\usepackage{adjustbox}
\usepackage{graphicx}
\usepackage{makecell}
\usepackage{lastpage}
\usepackage{tikz}
\usetikzlibrary{shapes,arrows.meta,positioning,calc,fit,backgrounds,patterns}

\definecolor{figBlue}{HTML}{2563eb}    %
\definecolor{figRed}{HTML}{dc2626}     %
\definecolor{figPurple}{HTML}{7c3aed}  %

\crefname{figure}{Fig.}{Figs.}
\Crefname{figure}{Fig.}{Figs.}

\crefname{table}{Table}{Tables}
\Crefname{table}{Table}{Tables}

\crefname{equation}{Eq.}{Eqs.}
\Crefname{equation}{Eq.}{Eqs.}

\usepackage[natbib=true,style=ieee,backend=biber,doi=true,url=false,isbn=false,maxbibnames=3,minbibnames=1]{biblatex}

\AtEveryBibitem{
    \clearfield{urlyear}
    \clearfield{urlmonth}
    \iffieldundef{doi}{}{\clearfield{url}}
    \clearfield{note}
    \clearlist{language}
}

\DefineBibliographyStrings{english}{
  url = \adddot\addspace Available ,
}

\DeclareCiteCommand{\textcitejournalclean}
  {}
  {\bibhyperref{\printnames{labelname}\addspace
   (\printfield{journaltitle}, \printfield{year})}}
  {\multicitedelim}
  {}

\graphicspath{{figures/}}
\newcommand{\SparTokens}{10.8}
\newcommand{\SparFeatures}{10.9}

\hypersetup{pdftitle={EMG Tokenization Preserves Movement Information at a Fraction of the Bitrate},pdfauthor={Raul C. S\^{i}mpetru, Amin Olamazadeh, Devon Rohlf, Vlad Cnejevici, and Alessandro Del Vecchio}}

\begin{document}
\raggedbottom
\title{EMG Tokenization Preserves Movement Information at a Fraction of the Bitrate}
\author{
  Raul C. S\^{i}mpetru \orcidlink{0000-0003-0455-0168},
  Amin Olamazadeh \orcidlink{0009-0002-6830-5559},
  Devon Rohlf \orcidlink{0000-0003-1601-0046},
  Vlad Cnejevici \orcidlink{0000-0003-2291-5941},
  and Alessandro Del Vecchio \orcidlink{0000-0002-7118-5554}
  \thanks{This work was supported in part by the European Research
    Council (ERC) through the GRASPAGAIN project under Grant 101118089
    and in part by the German Federal Ministry of Research, Technology
    and Space (BMFTR) through the BIONIK project under Grant 16SV9639,
    the EMGRoll project under Grant 16SV9306, and the NOE-EMY project
    under Grant 16SV9246, all to A. Del Vecchio. (Corresponding author: Alessandro Del Vecchio)}
  \thanks{The authors are with the Department of Artificial Intelligence in Biomedical Engineering, Friedrich-Alexander-Universität Erlangen-Nürnberg, 91052 Erlangen, Germany}
  \thanks{Code is available at \url{https://github.com/NsquaredLab/MyoJi}.}
}

\maketitle

\begin{abstract}
  \boldmath
  Wireless electromyography (EMG) interfaces must balance electrode
  coverage against transmission bandwidth. Here, we show that compact
  EMG tokens substantially reduce payload while retaining information
  for three-dimensional hand-pose decoding. We evaluate two causal
  tokenizers fitted without pose labels: an encoder pretrained to
  reconstruct EMG spectra and one based on SPAR, an engineered
  spectral--spatial representation of neighbouring electrodes. In
  personalized, within-session experiments with 320-channel EMG from
  nine participants, we compare tokenized and continuous features
  using matched decoder architectures and training procedures.
  Averaged across decoders, tokenization changed mean error by
  \(+0.01\)\,mm for SPAR (95\% confidence interval: \(-0.12\) to 0.14)
  and \(+0.45\)\,mm for the learned representation (0.10 to 0.82).
  Both increases met a 1\,mm non-inferiority criterion. Stored token
  arrays reduced payloads 41-fold relative to raw EMG and 36-fold
  relative to continuous SPAR features. On sampled windows of the public
  emg2pose benchmark's test partition, absolute
  differences in mean error between learned tokens and continuous
  features were below 0.05\,mm across 176 participants and in
  the subset of 20 participants unseen during training. Electrode
  coverage, calibration, and unseen recordings affected SPAR decoding
  more than quantization did. In a computational benchmark, optimized
  SPAR generated tokens for 63 channels within 20\,ms on a Cortex-M55
  microcontroller. These findings support tokenization as a compact
  interface between wearable sensing and movement decoding, allowing
  decoders to be updated while the wearable encoder and transmitted
  representation remain fixed.
\end{abstract}

\markboth{EMG Tokenization Preserves Movement Information at a Fraction of the Bitrate - S\^{i}mpetru et al. 2026}{}
\makeatletter
\def\ps@headings{%
  \def\@oddhead{\hbox{}\@IEEEheaderstyle\leftmark\hfil\thepage/\pageref{LastPage}}\relax
  \def\@evenhead{\@IEEEheaderstyle\thepage/\pageref{LastPage}\hfil\leftmark\hbox{}}\relax
  \let\@oddfoot\@empty
  \let\@evenfoot\@empty
}
\def\ps@IEEEtitlepagestyle{%
  \def\@oddhead{\hbox{}\@IEEEheaderstyle\leftmark\hfil\thepage/\pageref{LastPage}}\relax
  \def\@evenhead{\@IEEEheaderstyle\thepage/\pageref{LastPage}\hfil\leftmark\hbox{}}\relax
  \let\@oddfoot\@empty
  \let\@evenfoot\@empty
}
\makeatother
\pagestyle{headings}
\thispagestyle{IEEEtitlepagestyle}

\section{Introduction}\label{sec:introduction}
\IEEEPARstart{S}{urface} electromyography (sEMG) is a non-invasive method for measuring muscle activity \citep{merlettiEditorialSurfaceElectromyography2021} that is increasingly moving beyond laboratory settings, driven in part by its ease of use and growing commercial interest, from myoelectric prostheses \citep{farinaExtractionNeuralInformation2014} and wireless high-density amplifiers \citep{ceroneModularSmartWearable2019} to wristbands for computer input \citep{kaifoshGenericNoninvasiveNeuromotor2025,salterEmg2poseLargeDiverse2024}.

\begin{figure*}[!t]
  \centering
  \includegraphics[width=\linewidth]{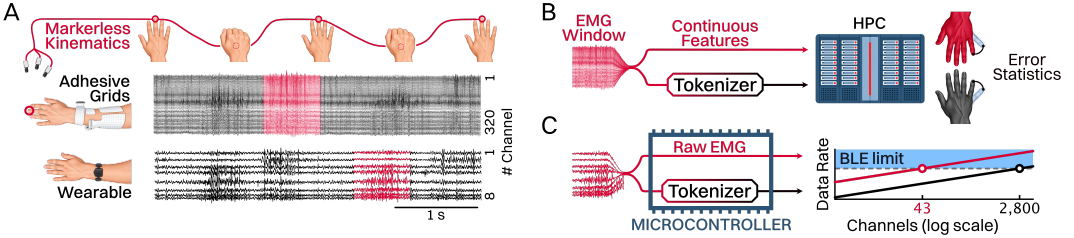}
  \caption{\textbf{From high-density electromyography (EMG) recordings to a transmittable representation.}
    \textbf{A:} Nine participants performed individual finger movements, grasping, and two- and three-finger pinches. Markerless hand tracking was recorded alongside surface EMG (sEMG) from 320 electrodes distributed across five grids. The lower band illustrates an eight-channel subset obtained by selecting every 40th channel from the 320-channel recording.
    \textbf{B:} Continuous EMG features and discrete tokens are used to train three-dimensional (3D) hand-pose decoders on a high-performance computing (HPC) system. Their decoding errors are compared using the same decoder architecture and training procedure within each pair.
    \textbf{C:} An assumed Bluetooth LE (BLE) 2M application throughput of 1.4\,Mbit/s \citep{bluetoothsigLEprimer} supports up to 42 channels of raw 16-bit EMG at 2,048\,Hz, and the 43rd exceeds the budget. Packed 10-bit tokens at 50\,Hz would support 2,800 channels. These are payload-only calculations before application packet headers, not measured transmission or computational capacities (Section~\ref{S-app:bandwidth}; Fig.~\ref{S-fig:operating-points}; Supplementary Data~1).}
  \label{fig:concept}
\end{figure*}

Wireless sEMG wearables \citep{simpetruMyoGesticEMGInterfacing2025,
  rolandinoHDEarrayDevelopmentValidation2024,ceroneModularSmartWearable2019,
  ceroneDesignValidationWireless2022}
trade electrode coverage against transmission bandwidth, even
though motor unit activity can differ between regions of a single
muscle \citep{hallerVoluntaryDissociationMotor2026}.
Compressing multichannel recordings could reduce bitrate while
retaining information useful for movement decoding
(Fig.~\ref{fig:concept}).

Electromyography (EMG) interfaces typically transmit either raw signals or continuous feature representations \citep{simpetruMyoGesticEMGInterfacing2025,simpetruProportionalSimultaneousRealTime2023,simpetruLearningHandModel2024,rahimiSimultaneousControlHuman2026,salterEmg2poseLargeDiverse2024}. An alternative is to encode EMG signals as discrete tokens that index vectors in a learned codebook. At the receiver, these vectors are retrieved and passed to the pose decoder. Vector quantization (VQ) has previously been explored for EMG signal reconstruction \citep{gronforsVectorQuantizationMethod2006}, while more recent approaches have introduced learned token representations for general time series \citep{talukderTOTEMTOkenizedTime2025} and biosignals \citep{avramidis2025biocodec,barmpas2026neurorvq}. In the context of hand-pose decoding, VQ-MyoPose uses vector quantization to encode ground-truth joint-angle trajectories and trains an EMG model to predict the resulting pose tokens \citep{lovecchio2025vqmyopose}. EMG tokenization separates wearable encoding from receiver-side task decoding.

We examine how converting continuous EMG features into discrete tokens affects hand-pose decoding accuracy, transmission bandwidth, memory requirements, and computational cost. We compare two approaches: a temporal encoder, shared across channels and pretrained to reconstruct EMG spectra, followed by vector quantization (Fig.~\ref{fig:learned-tokenizer}), and SPAR, which applies vector quantization to an engineered \textbf{Sp}ectral--Sp\textbf{A}tial \textbf{R}epresentation.\footnote{Fittingly, \emph{spar} is German for ``save''.} Both are fitted without pose labels, and separate models are then trained to predict three-dimensional (3D) hand kinematics. To assess the effect of quantization, we compare each tokenized representation with its continuous counterpart using the same decoder architecture and training procedure, with separate models fitted to each representation.
We evaluate both tokenizers on high-density EMG (HD-EMG) synchronized with 3D hand kinematics from nine participants \citep{simpetruLearningHandModel2024, osswaldTaskDependentMotorUnit2026}, using personalized decoders trained and tested within each participant's recording session. An adapted end-to-end waveform convolutional neural network (CNN) based on our previous work \citep{simpetruLearningHandModel2024, simpetruProportionalSimultaneousRealTime2023} provides a raw-EMG reference. The learned tokenizer is also evaluated on emg2pose \citep{salterEmg2poseLargeDiverse2024}, a public benchmark of 16-channel wrist EMG and motion-captured hand poses from 193 participants, including participants unseen during training. We also measure payload, memory, decoder cost (size and latency), and tokenizer execution time on a Cortex-M55 microcontroller (Section~\ref{S-app:contract}).

\section{Methods}
\label{sec:method}

\subsection{Dataset and acquisition}
We used our previously collected dataset of synchronized sEMG and hand-movement recordings from 13 healthy participants (11 men and 2 women; age \(25.9\pm2.8\)~years) \citep{simpetruLearningHandModel2024, simpetruInfluenceSpatiotemporalFiltering2024}. The nine participants analysed for decoding (participants 1--9) comprised seven men and two women. Eight were right-handed, one was left-handed, and one had a visible physiological tremor \citep{simpetruLearningHandModel2024}. The other four participants were used only for tokenizer fitting, never for decoding (Section~\ref{sec:experiments}).
The main decoding analysis used eight movement tasks of approximately 40\,s each, paced by a virtual hand on a monitor \citep{simpetruLearningHandModel2024}: flexion and extension of each digit at 0.5\,Hz, grasp opening and closing, and two- and three-finger pinching.
The experiments were approved by the ethics committee of Friedrich-Alexander-Universität Erlangen-Nürnberg (application no. 21-150-B) and conducted in accordance with the Declaration of Helsinki. All participants provided written informed consent.

Monopolar sEMG was acquired from 320 electrodes in five grids (OT Bioelettronica SRL, Turin, Italy): three \(8\times8\) grids around the forearm and two \(13\times5\) grids around the wrist, each with its final site unrecorded (64 electrodes) (\cref{fig:concept}A).
Signals were sampled at 2,048\,Hz with an EMG-Quattrocento amplifier (OT Bioelettronica).
Five synchronized cameras (DFK-37BUX287, The Imaging Source, Bremen, Germany) recorded the hand at 120\,Hz, and DeepLabCut \citep{mathisDeepLabCutMarkerlessPose2018} and Anipose \citep{karashchukAniposeToolkitRobust2021a} yielded 3D joint coordinates with a reported mean absolute localization error of 3\,mm relative to manual annotations \citep{cakiciGeneralizedFrameworkStudy2022}.

\subsection{Learned spectral tokenizer}
\begin{figure}[!t]
  \centering
  \includegraphics[width=\columnwidth]{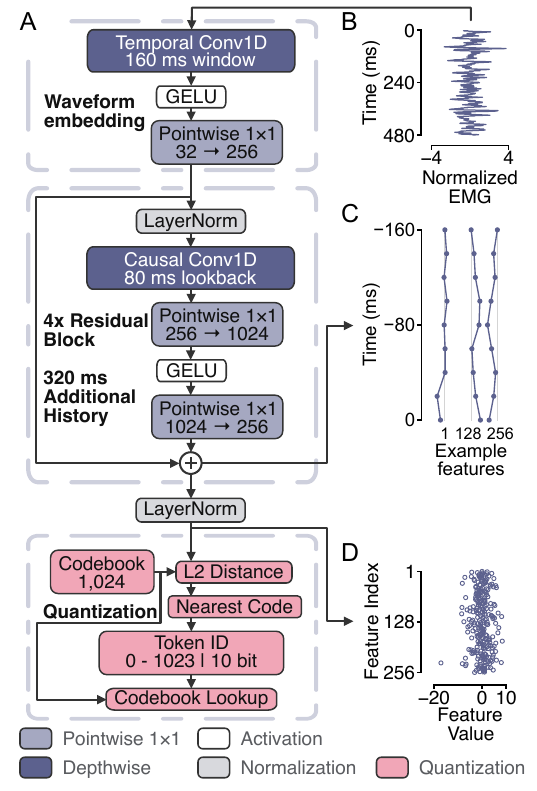}
  \caption{\textbf{Learned electromyography (EMG) tokenizer and intermediate representations.}
    \textbf{A:} A shared encoder processes 160\,ms EMG windows every 20\,ms. Four residual blocks extend the receptive field to 480\,ms, while a fixed codebook maps each 256-dimensional output to one of 1,024 tokens (10 bits). Layer dimensions such as \(32\rightarrow256\) denote the input and output feature dimensions. Conv1D denotes a one-dimensional convolution, GELU the Gaussian error linear unit, LayerNorm layer normalization, and L2 the Euclidean distance.
    \textbf{B:} Example normalized EMG waveform spanning 480\,ms (participant 1, electrode 1).
    \textbf{C:} Three of the 256 intermediate features (indices 1, 128, and 256) after the second residual block, shown at nine consecutive time points spaced 20\,ms apart.
    \textbf{D:} The resulting 256-dimensional feature vector, mapped to token identifier (ID) 93.
    Time zero in C corresponds to 480\,ms in B.}
  \label{fig:learned-tokenizer}
\end{figure}

The learned tokenizer is a custom causal implementation of the vector-quantized variational autoencoder (VQ-VAE) framework \citep{oord2017vqvae}, trained to reconstruct EMG spectra (Fig.~\ref{fig:learned-tokenizer}).
The same temporal encoder and codebook are applied to every EMG
channel independently, across electrode layouts
(Section~\ref{S-app:tokenizer}).

\emph{Encoding.}
EMG is resampled from 2,048 to 2,000\,Hz to match emg2pose, on which the encoder is also pretrained, and
normalized per channel using the median and scaled median absolute deviation
of the first 500\,ms of each 3\,s input window, with clipping at
\(\pm12\) scale units. Each window is therefore calibrated from its
own first 500\,ms, independently of earlier data.
Section~\ref{sec:sensitivity} evaluates calibration at recording
onset for SPAR. The encoder processes 160\,ms windows
at 20\,ms intervals, and four causal residual blocks give a 480\,ms
receptive field (Fig.~\ref{fig:learned-tokenizer}A).
We call each 20\,ms step, at which every channel emits one token, an update.
After layer normalization, each 256-dimensional output is mapped
to the nearest of 1,024 codebook vectors by squared Euclidean
distance. The resulting index requires 10 bits when packed.
At the receiver, these vectors are retrieved, and the pose decoder
estimates 3D hand kinematics from their history across channels.
The first token with a complete receptive field is available 980\,ms
after window onset (500\,ms calibration plus 480\,ms context), excluding computation, and subsequent tokens follow at 50\,Hz (Section~\ref{S-app:tokenizer}).

\emph{Pretraining.}
The tokenizer is pretrained without kinematic labels on our
high-density recordings and emg2pose
\citep{salterEmg2poseLargeDiverse2024}, with the data separation specified in Section~\ref{S-app:data}. The encoder is trained
to reconstruct, under a smooth \(L_1\) loss, the log-power spectrum
of each mean-removed, Hann-tapered window in 64 nonzero-frequency bins up to 400\,Hz. Training uses a
straight-through gradient estimator \citep{oord2017vqvae}, a
commitment loss, a penalty that discourages use of only a small subset of codebook entries, and exponential-moving-average
codebook updates (Section~\ref{S-app:tokenizer}).

\subsection{SPAR: spectral--spatial features and tokenization}

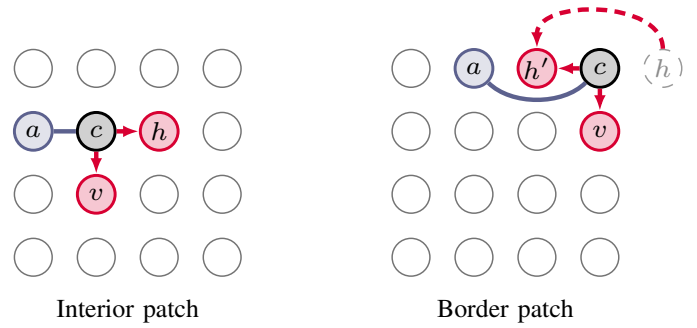
\begin{figure}[!t]
  \centering
  \definecolor{sparPatchBlue}{HTML}{5c618e}
  \definecolor{sparPatchRed}{HTML}{d80032}
  \resizebox{\columnwidth}{!}{%
    \begin{tikzpicture}[x=6mm,y=6mm,
        site/.style={circle,draw=black!55,fill=white,minimum size=3.6mm,inner sep=0pt,font=\scriptsize},
        centre/.style={site,fill=black!18,draw=black,thick},
        kept/.style={site,fill=sparPatchRed!22,draw=sparPatchRed,thick},
        aux/.style={site,fill=sparPatchBlue!18,draw=sparPatchBlue,thick},
        edge/.style={-{Latex[length=1.5mm]},very thick,sparPatchRed},
        sub/.style={-{Latex[length=1.5mm]},very thick,sparPatchRed,densely dashed},
        auxe/.style={very thick,sparPatchBlue},
        ghost/.style={site,draw=black!40,dashed,text=black!45},
        lab/.style={font=\scriptsize,inner sep=1pt}]
      \foreach \r in {0,...,3} \foreach \c in {0,...,3} \node[site] at (\c,-\r) {};
      \node[aux] (ia) at (0,-1) {\(a\)};
      \node[kept] (ih) at (2,-1) {\(h\)};
      \node[kept] (iv) at (1,-2) {\(v\)};
      \node[centre] (ic) at (1,-1) {\(c\)};
      \draw[edge] (ic) -- (ih); \draw[edge] (ic) -- (iv); \draw[auxe] (ic) -- (ia);
      \node[lab,anchor=north,align=center] at (1.5,-3.6) {Interior patch};
      \begin{scope}[xshift=36mm]
        \foreach \r in {0,...,3} \foreach \c in {0,...,3} \node[site] at (\c,-\r) {};
        \node[aux] (ba) at (1,0) {\(a\)};
        \node[kept] (bh) at (2,0) {\(h'\)};
        \node[kept] (bv) at (3,-1) {\(v\)};
        \node[centre] (bc) at (3,0) {\(c\)};
        \node[ghost] (bm) at (4,0) {\(h\)};
        \draw[edge] (bc) -- (bh); \draw[edge] (bc) -- (bv); \draw[sub] (bm) to[out=90,in=90] (bh);
        \draw[auxe] (bc) to[out=-135,in=-45,looseness=.85] (ba);
        \node[lab,anchor=north,align=center] at (1.5,-3.6) {Border patch};
      \end{scope}
    \end{tikzpicture}}
  \caption{\textbf{Neighbourhoods and boundary handling of the spectral--spatial representation.}
    Grey marks the centre electrode, red the horizontal and vertical neighbours, and blue the auxiliary electrode. At a boundary, an electrode on the opposite side of the centre replaces the missing neighbour (dashed arrow), and its coupling is set to zero after the matrix logarithm. Each band contributes the centre's diagonal entry and the real horizontal and vertical couplings.}

  \label{fig:spar-patch}
\end{figure}

SPAR directly computes local spectral--spatial features describing each
electrode's spectrum and coupling to neighbouring electrodes, as an
alternative to learned per-channel spectral features
(Section~\ref{S-app:spar}, Fig.~\ref{fig:spar-patch}).

\emph{Local patch.}
Each electrode defines a four-electrode patch within its grid:
the centre, horizontal and vertical neighbours, and an auxiliary
electrode. The auxiliary electrode contributes to the cross-spectral matrix
and therefore affects its logarithm, but no features are extracted
directly from its entries. At grid boundaries, an electrode on the
opposite side of the centre replaces the missing neighbour in the
patch. The substitute still enters the cross-spectral matrix, but its
coupling features are set to zero after the matrix logarithm, so that no
feature describes an absent neighbour
(Section~\ref{S-app:spar}, Fig.~\ref{fig:spar-patch}).

\emph{Spectral matrices.}
Each causal 160\,ms window is divided into seven overlapping
64\,ms Hann segments, spaced 16\,ms apart. Their Fourier
coefficients are used to compute a \(4\times4\) cross-spectral
matrix in six frequency bands (20--40, 40--60, 60--100,
100--160, 160--250, and 250--500\,Hz):
\begin{equation}
  \mathbf{S}_{c,b,t}
  = \frac{1}{7}\sum_{s=1}^{7}\sum_{f\in b}
  w_{f,t}\,\mathbf{X}_{c,s,f,t}\,
  \mathbf{X}_{c,s,f,t}^{\mathsf{H}},
  \label{eq:spar-matrix}
\end{equation}
where \(\mathbf{X}_{c,s,f,t}\) contains the Fourier coefficients
of the four signals and \(w_{f,t}\) weights each frequency bin
according to its relative contribution to the band's power,
computed across all channels. A small diagonal regularization and an
eigenvalue floor keep the matrix logarithm defined, even for silent
patches, yielding \(\mathbf{L}_{c,b,t}
= \log\mathbf{S}^{\mathrm{reg}}_{c,b,t}\).
Related electroencephalography (EEG) methods also combine cross-spectral matrices with
logarithmic mappings \citep{li2022harmnqeeg,lopezNaranjo2024riemannian}.

\emph{Features.}
From each logarithmic matrix, we extract three features:
the diagonal entry of the centre electrode and the real
parts of its couplings with the right and lower
neighbours. Across six frequency bands, this produces
18 features per electrode. Features of missing neighbours are fixed at zero (structural zeros).

\emph{Tokenization.}
The 18 feature components are standardized using statistics
from each participant's training windows, excluding the structural
zeros. For each participant, \(k\)-means
clustering fits a codebook of 1,024 centroids to sampled
training feature vectors without kinematic labels. The codebook
is shared across electrodes, but not across participants
(Section~\ref{S-app:spar}).

Each feature vector is replaced by the index of its nearest
centroid, requiring 10 bits when packed. The receiver restores
the corresponding centroid and re-applies the structural zeros.
The codebook remains fixed during pose-decoder training.
Continuous SPAR uses the same standardized features without
quantization. We also evaluate a 64-centroid codebook,
requiring 6 bits per token (Section~\ref{sec:manifold};
Section~\ref{S-app:spar}). SPAR was
defined for the monopolar electrode grids of our HD-EMG recordings and
was not adapted to the bipolar emg2pose wristband.

\subsection{Evaluation protocol}
\label{sec:experiments}
Each of the nine participants contributes 64 training,
16 validation, and 16 test windows of 3\,s each across
eight movement tasks. Windows are divided chronologically
within each task recording, with at least 1\,s separating
the training, validation, and test blocks. Decoders predict 20 wrist-relative 3D hand landmarks,
and for the pose decoders only the final second of each window contributes to
scoring: 50 complete hand-pose predictions at 50\,Hz.
The two test windows per movement therefore contribute
two scored seconds (Fig.~\ref{fig:decoding};
Section~\ref{S-app:training}).

The learned tokenizer was pretrained without kinematic labels
using data from all 13 participants, excluding the validation
and test windows of the nine participants used for decoding. SPAR
feature standardization and codebook fitting used only each
participant's training windows. These experiments assess
personalized, within-session decoding, not generalization
to new participants or sessions. To examine whether small quantization
effects persist when tokenizer fitting excludes decoding
participants, a separate analysis fits tokenizers only on the
four participants outside the decoding cohort
(Fig.~\ref{S-fig:quantization-transfer}A; Section~\ref{S-app:exposure};
Supplementary Data~4).

The learned representation is also evaluated on a deterministic
subset of windows from the official partitions of the wrist-EMG
emg2pose benchmark \citep{salterEmg2poseLargeDiverse2024}. Its test
subset covers 176 participants, including 20 unseen during tokenizer
pretraining and decoder training (Sections~\ref{S-app:codec}
and~\ref{S-app:emg2pose}; Figs.~\ref{S-fig:emg2pose-cohorts}
and~\ref{S-fig:emg2pose-comparison}; Supplementary Data~4--7).

\subsection{Decoders and baseline comparisons}
Sections~\ref{S-app:decoders} and~\ref{S-app:codec} specify the
architecture and training of every decoder. Each decoder is fitted separately to
tokens and to their continuous counterpart with the same architecture and procedure, so these pairs measure the cost of
tokenization (Fig.~\ref{fig:dpose-main}C). For the learned
representation, the continuous input is the same encoder's
256-dimensional output before quantization, not the output of a
separately trained continuous encoder. For SPAR, it is the standardized features
without quantization (Section~\ref{S-app:training}).

We compare several decoder architectures to assess whether the
effect of tokenization depends on decoder design.

Raw-EMG waveform CNNs provide a reference for decoding performance
without tokenization. We evaluate both a CNN adapted from our earlier
work \citep{simpetruProportionalSimultaneousRealTime2023,
  simpetruLearningHandModel2024} and the original CNN published in IEEE Transactions on Biomedical
Engineering (TBME CNN) \citep{simpetruLearningHandModel2024}, retrained
with its published preprocessing and an adapted training recipe
(Section~\ref{S-app:decoders}).

The causal temporal convolutional network (TCN) and Transformer decoders
\citep{bai2018tcn,vaswani2017attention} and a third waveform CNN matched
to them each have approximately 7.5 million parameters and use the same
training budget. This allows us to compare input representations
while controlling for model size and training resources. Results
for this matched waveform CNN are reported in Supplementary Data~2.

The learned tokenizer encodes each channel independently. To test
whether a small decoder can recover spatial patterns, our proposed
decoder, a compact spatial CNN + TCN (about 0.7 million parameters),
applies convolutions to a five-array by 64-electrode arrangement in
acquisition order before temporal modelling.

Tokenization is often motivated by compatibility with large
pretrained sequence models. The Qwen3 language model
\citep{qwen3technicalreport}, kept frozen and receiving codebook vectors
through a trained adapter, tests whether language-model pretraining
helps EMG decoding. Two controls replace the pretrained weights with
random frozen weights or remove the backbone so that the adapter feeds
the pose output directly (direct pose readout in Fig.~\ref{fig:dpose-main}A).

Tokens also allow the
decoder to change without changing the wearable: frozen TabPFN
\citep{hollmann2025tabpfn,priorlabs2026tabpfn35}, a pretrained
tabular model, decodes SPAR from labelled examples supplied as
context, without training its weights on the task. The paired comparisons
isolate the effect of quantization, whereas comparisons between decoders
evaluate complete pipelines.

\subsection{Communication and computational requirements}
The token payload rate is the product of channel count, update
rate, and bits per token. At 50\,Hz, 320 channels require 20\,kB/s with packed
10-bit indices or 32\,kB/s in stored 16-bit arrays. Application
headers and codebook memory are accounted for separately
(Section~\ref{S-app:bandwidth}; Supplementary Data~1).
Wireless capacity assumes 1.4\,Mbit/s Bluetooth LE 2M application
throughput \citep{bluetoothsigLEprimer}, the ideal case of continuous
one-way transfer of maximum-length (251-byte) packets
\citep[Vol.~6, Part~B, Sec.~2.4]{bluetooth_sig_bluetooth_2026} without
retransmissions, including lower-layer overhead but excluding our
application headers. Practical links achieve less.
Decoder cost is measured as model size and 32-bit floating-point (FP32) latency on an RTX 4090 graphics card (NVIDIA, Santa Clara, CA, USA).
Microcontroller timing on a NUCLEO-N657X0-Q board (STMicroelectronics, Geneva, Switzerland) tests the 20\,ms token-generation budget, and separate checks test whether the optimized code reproduces the reference tokens and how pose accuracy changes
at reduced electrode coverage (Section~\ref{S-app:contract}).

\subsection{Representation visualization and pose retrieval}
\label{sec:representation}
To visualize how tokens evolve during movement, Fig.~\ref{fig:token-pose}A--C
uses participant 1's eight movement recordings. At each update,
the 320 SPAR tokens are replaced by their 18-dimensional codebook
vectors. We concatenate the current representation with those
120, 240, 360, and 480\,ms earlier, then fit a three-dimensional
Uniform Manifold Approximation and Projection (UMAP)
\citep{mcinnesUMAPUniformManifold2018a} using cosine distance.
Movement labels and hand poses are not inputs to the UMAP fit.

The two-dimensional display is constructed from this embedding.
Peaks in the measured mean fingertip-to-wrist distance define repetition cycles and
their normalized phase. For each movement, we align and average
its cycles, project its points onto the plane of the mean
trajectory, and reposition the group to avoid overlap. Thus,
movement labels and pose-derived phase influence the display, and distances between groups are not interpretable. Panel C summarizes
repetitions using phase bins and smoothing, followed by periodic
cubic interpolation that explicitly closes the curves. Exact
embedding and display settings are given in
Section~\ref{S-app:umap}.

Pose retrieval tests whether a representation carries pose information without a trained decoder, and direction classification tests whether it distinguishes opening from closing at the same hand closure.
Both use all nine participants and one SPAR representation per update,
each computed from the preceding 160\,ms of EMG, without UMAP or
concatenated history. For pose retrieval,
queries are eligible updates whose complete 160\,ms analysis window
lies within a test window, including updates before the final second
scored for the pose decoders. We exclude the query's own
and adjacent cycles, find the closest update in each remaining
repetition cycle outside the training windows, and average the
measured poses from the five closest cycles. Matching uses Euclidean distance
between feature or codebook vectors, or Hamming distance between
token identifiers (IDs). Direction classification uses every in-cycle update as a query and
matches it to updates at similar hand closure in other cycles (Section~\ref{S-app:manifold}).

\subsection{Robustness analyses}
Simulated noise and electrode loss test whether quantization
adds sensitivity to degraded inputs in decoders that are not
retrained for them (Section~\ref{S-app:corruption}). We first changed the
calibration procedure at evaluation without refitting the pipelines,
testing sensitivity to a mismatch between training and evaluation.
We then refitted the pipelines under the new calibration procedure to
assess performance when training and evaluation used the same procedure.
Unseen recordings, not used to fit the evaluated SPAR pipelines,
assess performance
under changed recording and movement conditions (Section~\ref{S-app:replays}).

\subsection{Outcomes and statistical analysis}
\label{sec:stats}
The primary outcome is mean Euclidean distance between
predicted and reference landmarks, in millimetres. Errors
are averaged across valid predictions and evaluated landmarks
within each participant, then across participants with equal
weight. We evaluate 20 landmarks for HD-EMG and 16 for emg2pose (Section~\ref{S-app:codec}).

\begin{figure*}[!t]
  \centering
  \includegraphics[width=0.98\linewidth]{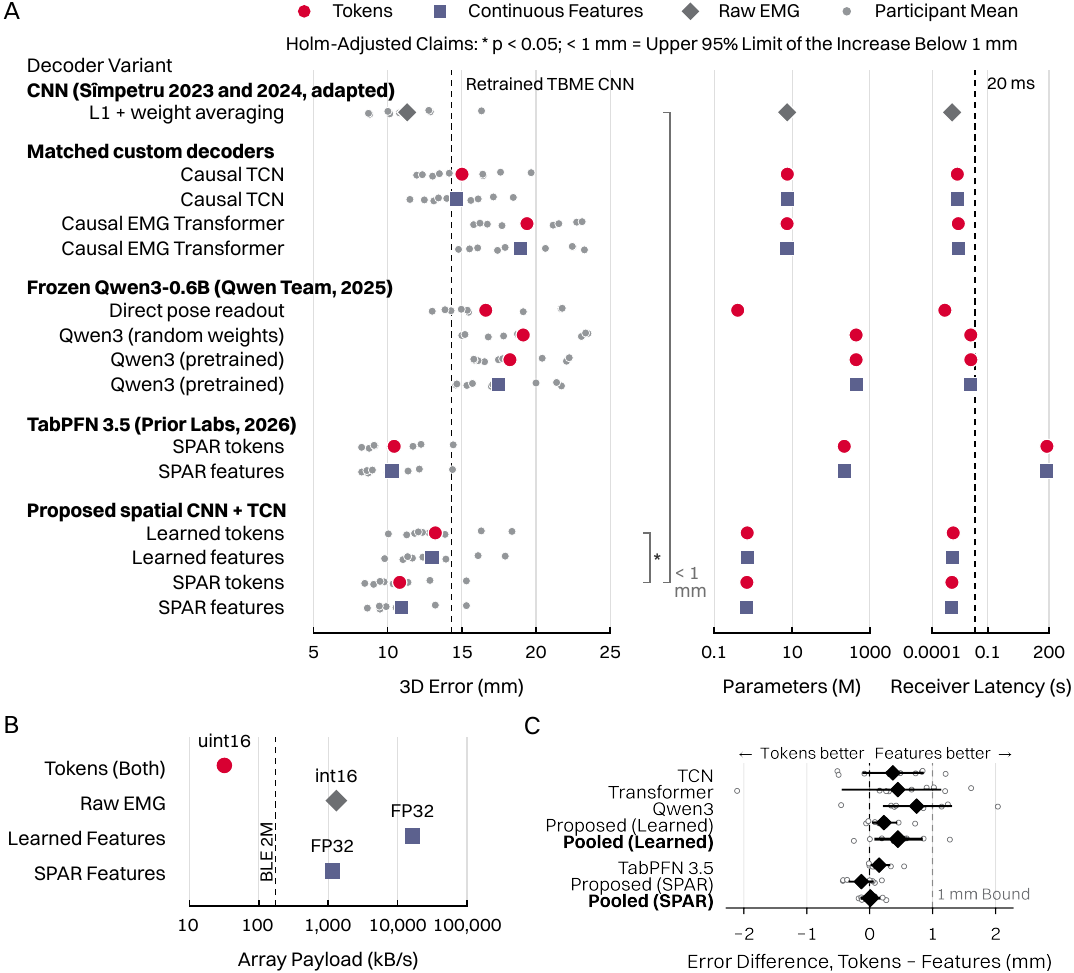}
  \caption{\textbf{Hand-pose decoding accuracy, computational cost, and quantization effects.}
    \textbf{A:} Unfiltered three-dimensional (3D) error over 20 wrist-relative landmarks, decoder parameters including frozen backbones, and 32-bit floating-point (FP32) decoder latency. Qwen3 \citep{qwen3technicalreport} and TabPFN \citep{hollmann2025tabpfn,priorlabs2026tabpfn35} are frozen pretrained decoders, and TCN is a temporal convolutional network. Grey points are participant means over three seeds, and coloured symbols group means. Latency is the mean and standard deviation per complete-pose prediction (batch size one, RTX 4090), excluding feature extraction and tokenization (TabPFN also includes host-processor work). Dashed lines mark the 20\,ms update interval and the retrained convolutional neural network (CNN) from IEEE Transactions on Biomedical Engineering (TBME CNN) \citep{simpetruLearningHandModel2024}. Brackets compare the spectral--spatial representation (SPAR) with learned tokens in the proposed decoder (*, \(p=0.016\)) and SPAR tokens with the adapted CNN (< 1\,mm, \(p=0.016\)).
    \textbf{B:} Stored-array payload for 320 channels before packet headers. The dashed line marks the assumed Bluetooth LE (BLE) 2M throughput of 175\,kB/s \citep{bluetoothsigLEprimer}. Labels give stored types: unsigned (uint16) and signed (int16) 16-bit integers for tokens and raw electromyography (EMG), and FP32 for features.
    \textbf{C:} Token error minus continuous-feature error. Circles are participants, diamonds are means, and bars are 95\% confidence intervals. Bold Pooled rows average the differences across decoders within each tokenizer, and both upper 95\% confidence limits lie below the dashed 1\,mm bound (\(p=0.016\) each). The four \(p\)-values in A and C are from exact paired sign-flip permutation tests, Holm-adjusted together (Section~\ref{sec:stats}).}
  \label{fig:dpose-main}
\end{figure*}

\begin{figure*}[!t]
  \centering
  \includegraphics[width=\linewidth]{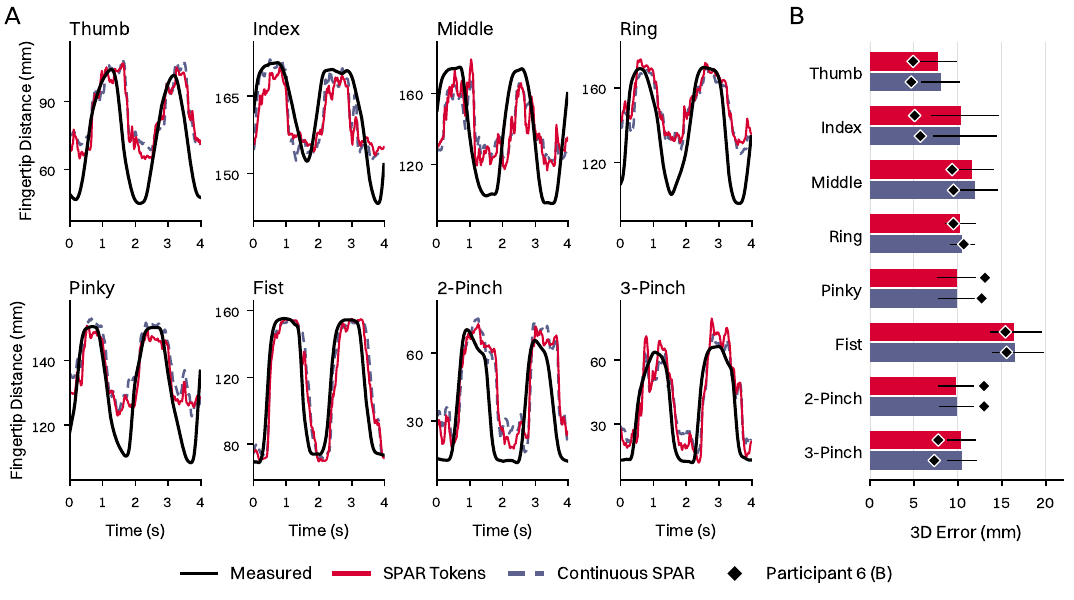}
  \caption{\textbf{Decoded hand movements and movement-specific errors.}
    \textbf{A:} Example predictions from participant 6. Traces span 4\,s, and scoring uses the final second of each predefined test window. Black indicates measured movement, red spectral--spatial representation (SPAR) tokens, and dashed blue continuous SPAR. Predictions are smoothed with a causal One Euro filter for display only. Traces show the thumb-tip-to-pinky-base distance for the thumb, the fingertip-to-wrist distance for the other digits (its mean over fingertips for the fist), and the thumb-to-fingertip distance for the pinches (averaged over index and middle fingers for the three-finger pinch). Section~\ref{S-app:noise} describes the predictions shown between scored seconds and the filter.
    \textbf{B:} Unfiltered three-dimensional (3D) error across 20 landmarks for each movement, averaged over nine participants and three seeds. Bars show group means with 95\% confidence intervals, and diamonds identify participant 6. Supplementary Data~2 reports the paired comparisons.}
  \label{fig:decoding}
\end{figure*}

Quantization effects are participant-level paired differences:
token error minus continuous-feature error. Negative values
favour tokens. For HD-EMG, each participant's error is
averaged across the eight tasks and three decoder seeds before
differencing (Fig.~\ref{fig:dpose-main}C).
Movement-specific SPAR comparisons assess whether these averages
conceal differences across tasks (Fig.~\ref{fig:decoding}).

Participants are the units of statistical analysis.
We test whether tokens increase error by less than a 1\,mm
engineering bound (a non-inferiority test \citep{piaggio2012consort}),
a third of the 3\,mm localization error of the reference
kinematics (Section~\ref{S-app:statistics}). We apply Holm correction \citep{holm1979simple} across four
comparisons: an increase below 1\,mm for SPAR and for learned
tokens, each averaged within participant over its decoders; SPAR versus learned
tokens with the proposed decoder; and non-inferiority of SPAR
tokens to the adapted CNN with the 1\,mm bound. Each uses an exact paired permutation (sign-flip)
test over all 512 sign patterns \citep{winkler2014permutation}, and statistical significance is assessed using the
Holm-adjusted \(p\)-values. One-sided non-inferiority \(p\)-values
are doubled before Holm correction, so the smallest attainable
Holm-adjusted \(p\)-value is 0.016. HD-EMG comparisons without a
specified secondary family are reported as estimates with exact
pointwise 95\% confidence intervals. The
procedures assume independent participants and symmetric
differences. Secondary analyses are Holm-corrected within their own
families (Sections~\ref{S-app:statistics}, \ref{S-app:noise},
and~\ref{S-app:n6}).

\section{Results}
\label{sec:results}

\subsection{Averaged across decoders, quantization adds less than a millimetre of error}
\emph{Tokenization.}
We assessed the accuracy cost of tokenization using paired
token--feature comparisons. Averaged across decoders, tokenization
changed error by \(+0.01\)\,mm for SPAR
(95\% confidence interval: \(-0.12\) to \(+0.14\)\,mm)
and \(+0.45\)\,mm for the learned representation
(\(+0.10\) to \(+0.82\)\,mm).
Both increases were below the 1\,mm bound
(Holm-adjusted \(p=0.016\) for each;
Fig.~\ref{fig:dpose-main}C).
However, the bound could not be confirmed for every decoder
individually: for the learned representation, the upper 95\% confidence
limits reached 1.12\,mm with the Transformer and 1.29\,mm with Qwen3 (Supplementary Data~2).
For SPAR, decoding error varied across movements, but
for every movement the pointwise upper 95\% confidence limit of the
token-minus-continuous difference was at or below 0.59\,mm, below the 1\,mm bound (Fig.~\ref{fig:decoding}B).
These differences also remained small when tokenizers
were fitted exclusively on participants outside the
nine-person decoding cohort. The pose decoders were still
trained separately for each decoding participant, so this
analysis tests tokenizer transfer rather than generalization
of the full decoding pipeline to unseen participants
(Fig.~\ref{S-fig:quantization-transfer}A;
Section~\ref{S-app:exposure}).

\emph{emg2pose benchmark.}
On emg2pose, learned tokens and continuous features gave mean
errors of 28.00 and 28.04\,mm across 176 test participants and
three independently pretrained encoders, a token-minus-continuous
difference of \(-0.04\)\,mm (95\% confidence interval \(-0.16\) to 0.07).
The difference was \(+0.02\)\,mm (\(-0.16\) to 0.21) in the
20 participants unseen during tokenizer pretraining and decoder
training, and \(+0.04\)\,mm (\(-0.49\) to 0.56) in held-out
same-day sessions from ten participants
(Fig.~\ref{S-fig:emg2pose-cohorts}B; Section~\ref{S-app:emg2pose};
Supplementary Data~4 and~6). These errors refer to the benchmark's
generic hand model and are not directly comparable with the
HD-EMG errors. Holding each model's final prediction before the scored
interval fixed throughout that interval gave 29.65\,mm for learned tokens
and 29.63\,mm for continuous features, and velocity \(R^2\) was near
zero for both representations (Section~\ref{S-app:motion}).

\emph{Complete-pipeline accuracy.}
Representation and decoder choice produced larger differences
in error than tokenization alone. The proposed spatial CNN + TCN
achieved \SparTokens{}\,mm with SPAR tokens and
\SparFeatures{}\,mm with continuous SPAR features.
With this decoder, SPAR tokens gave 2.4\,mm lower error than
learned tokens (95\% confidence interval: 1.7 to 3.1\,mm;
Holm-adjusted \(p=0.016\)). Within SPAR, tokens of each
electrode's own log band power gave 2.5\,mm higher error than
full SPAR tokens (95\% confidence interval: 1.6 to 3.4\,mm), and tokens of the matrix-logarithm
diagonal alone, without couplings, 0.3\,mm higher (0.1 to 0.5\,mm; Holm-adjusted
\(p=0.008\) each; Section~\ref{S-app:ablation}). The SPAR-token pipeline also met the 1\,mm non-inferiority
criterion relative to the adapted waveform CNN operating on raw EMG.
With learned-token inputs, the proposed decoder had lower
mean error than the matched TCN, matched Transformer, and
frozen Qwen3. For Qwen3, pretrained weights gave lower mean
error than random weights, but still higher error than the
control without the backbone
(Fig.~\ref{fig:dpose-main}A; Supplementary Data~2 and~3).

\emph{Decoder cost.}
The proposed decoder combined lower mean error with fewer
parameters than the matched temporal models. The waveform,
temporal, and Qwen3 decoders all had mean inference latencies
below the 20\,ms update interval. TabPFN reduced error by a
further 0.37\,mm relative to the proposed SPAR-token decoder,
but required more than two minutes per complete-pose prediction.
TabPFN timing includes processing on the central processing unit, and every prediction
reprocesses the labelled training context
(Fig.~\ref{fig:dpose-main}A; Supplementary Data~3).

\begin{figure*}[!t]
  \centering
  \includegraphics[width=1\linewidth]{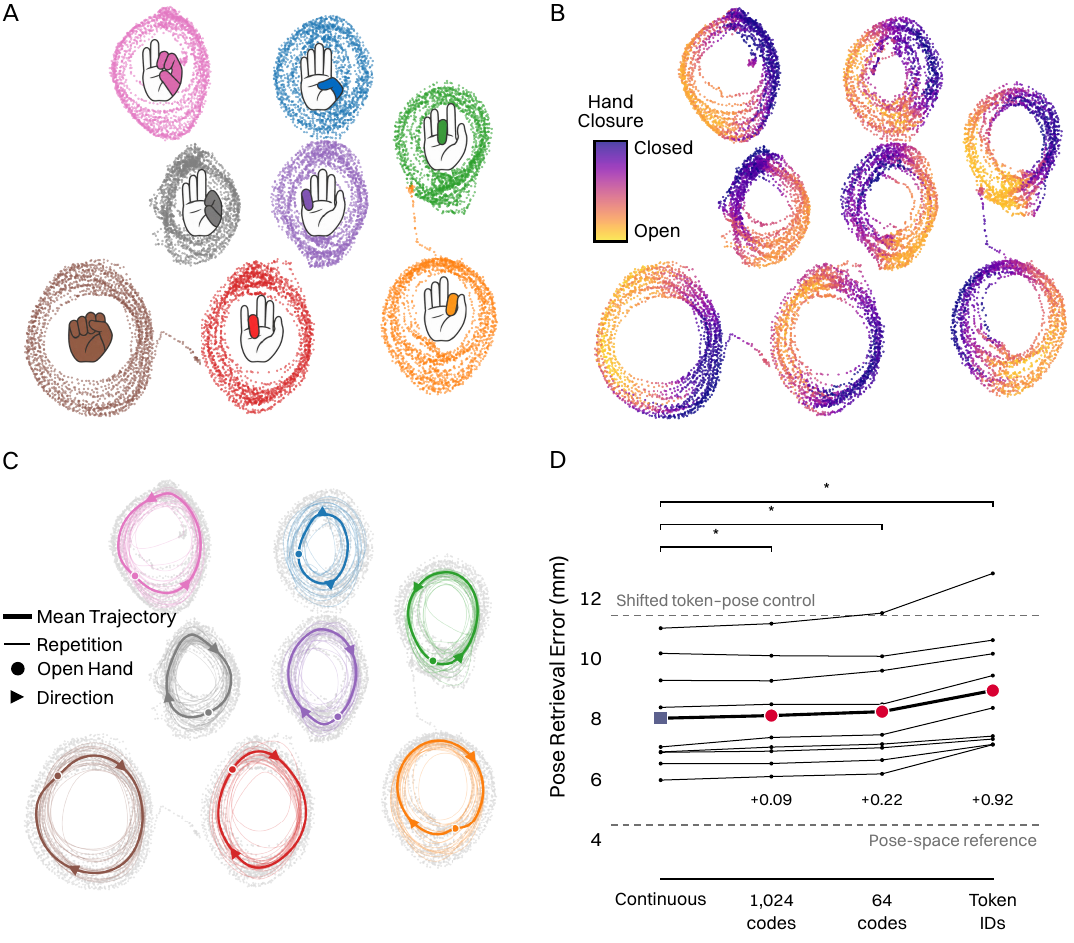}
  \caption{\textbf{Hand-pose structure in spectral--spatial representation (SPAR) token space.}
    \textbf{A--C:} One three-dimensional Uniform Manifold Approximation and Projection (UMAP) of participant 1's token histories across eight movements. Each movement is projected separately onto its mean-trajectory plane and repositioned for display (Section~\ref{sec:representation}).
    Figure~\ref{S-fig:token-history} shows how the projected trajectories change as token history increases from a single update to 960\,ms.
    \textbf{A:} Movement identity, with icons indicating the active fingers.
    \textbf{B:} Hand closure normalized separately for each movement.
    \textbf{C:} Phase-aligned, smoothed repetitions (thin) and their mean trajectory (thick), with cycle starts near the open-hand pose (dots) and traversal direction (arrows).
    \textbf{D:} Pose retrieval for single updates in the test windows, matched against repetitions outside the training windows, using continuous SPAR, dequantized tokens, or token identifiers (IDs) in the original representation space. Thin lines show the nine participants and the thick line their mean. Dashed lines are descriptive references: disrupted token--pose alignment (11.3\,mm, a mean of participant medians, for which 1,024-code tokens give 7.3\,mm) and matching using measured poses (4.4\,mm). The curves and the pose-space reference are phase-balanced means. Asterisks indicate increases over continuous SPAR (exact paired permutation (sign-flip) tests, Holm-adjusted across three comparisons; \(p=0.043\), \(0.016\), and \(0.012\)). Retrieval, display-selection rules, and reference construction are detailed in Section~\ref{S-app:manifold}.}
  \label{fig:token-pose}
\end{figure*}

\subsection{SPAR tokens retain hand pose across repetitions}
\label{sec:manifold}
A trained decoder can adapt to changes introduced by
quantization. We therefore also tested whether tokens
preserved pose-related similarity using a simple retrieval
analysis: for each update, we found similar updates in
other repetitions and averaged their recorded poses.
Queries came from the test windows and matches from
repetitions outside the training windows. When similarity
was measured using the codebook vectors identified by the
tokens, pose-retrieval error increased by 0.09\,mm with
1,024 codes and 0.22\,mm with 64 codes relative to
continuous SPAR (Fig.~\ref{fig:token-pose}D). Matching token
IDs directly with Hamming distance instead increased error
by 0.92\,mm. For the 1,024-code tokens, codebook-vector matching and Hamming
matching used the same transmitted token IDs and differed only in
whether matching accounted for distances between codebook vectors
(Section~\ref{S-app:manifold}).

\begin{figure*}[!t]
  \centering
  \includegraphics[width=\linewidth]{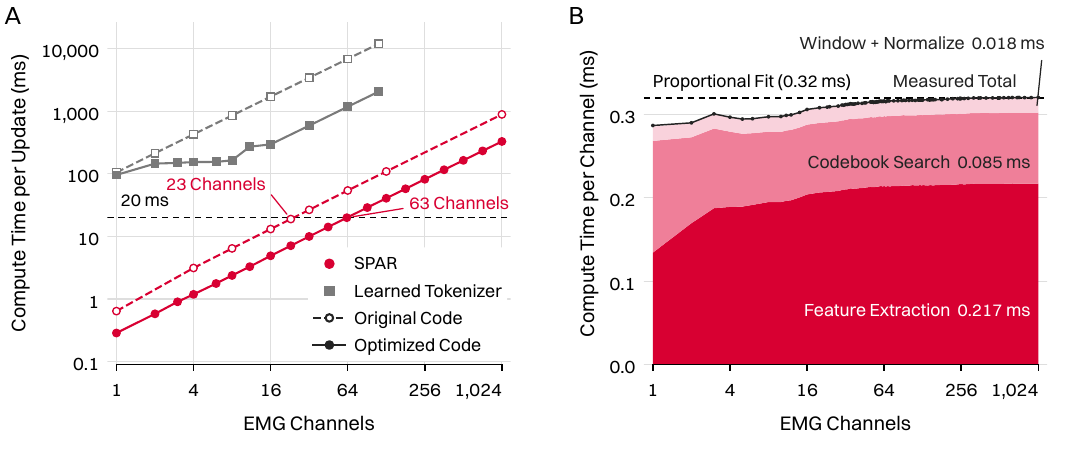}
  \caption{\textbf{Computational performance of electromyography (EMG) tokenization on a Cortex-M55 microcontroller.}
    The spectral--spatial representation (SPAR) implementations were tested at 600\,MHz with synthetic 2\,kHz EMG. Timing includes windowing, normalization, feature extraction, codebook search, and token packing, but excludes acquisition, radio transmission, and pose decoding.
    \textbf{A:} Median compute time per update over 16 timed updates, after 16 untimed warm-up updates. Open symbols and dashed lines show the original C implementations, and filled symbols and solid lines the optimized ones. Every timed update at the marked SPAR channel limits met the 20\,ms interval.
    \textbf{B:} Optimized SPAR time per channel by processing stage, with labels giving values at 320 channels. The dashed line shows time per channel from a linear fit of total time against channel count through the origin (Section~\ref{S-app:n6}).}
  \label{fig:stm32n6}
\end{figure*}

We also examined how tokens represented movement over time.
For participant~1, hand closure varied along the visualized
token-history trajectories (Fig.~\ref{fig:token-pose}A--C),
complementing our previous use of dimensionality reduction
to identify spared motor dimensions
\citep{simpetruIdentificationSparedProportionally2024}.

To test whether tokens retained movement-direction
information beyond hand closure, we separately classified
opening versus closing at matched hand closure in a descriptive
analysis that uses every in-cycle update as a query and the
original representations rather than the plotted coordinates. Accuracy was 70\% with dequantized tokens
and 71\% with continuous SPAR, compared with 55\% for the
majority-direction control (Section~\ref{S-app:manifold}).

\subsection{Communication cost and embedded feasibility}
At 320 channels, stored token arrays required 32\,kB per second
of recording, a 41-fold reduction relative to raw EMG and a 36-fold
reduction relative to continuous SPAR features. Ideal 10-bit packing
would reduce this to 20\,kB/s (Fig.~\ref{fig:dpose-main}B;
Supplementary Data~1). The learned encoder's continuous output,
256 FP32 values per channel and update, was 512 times larger
than its stored tokens. We benchmarked token generation to assess whether it fits within the
update interval.
On the Cortex-M55, optimization increased SPAR's tested
capacity from 23 to 63 channels within the 20\,ms interval
(Fig.~\ref{fig:stm32n6}). The optimized learned tokenizer
required 96\,ms for one channel and did not meet this
budget in the tested implementation. Run on a general-purpose computer, the optimized SPAR code reproduced
over 99.9\% of the reference implementation's token assignments with negligible change
in mean pose error (Section~\ref{S-app:n6}).

In a separate offline analysis, SPAR codebooks and decoders
refitted on the first 63 electrodes (the first forearm array
without its final corner electrode) achieved 12.5\,mm with
tokens, compared with 10.8\,mm using 320 electrodes. The assumed Bluetooth LE 2M throughput supports only 42 channels of
raw 16-bit EMG. The adapted CNN refitted on the first
42 electrodes had a mean error of 15.0\,mm, compared with 12.5\,mm
for the 63-channel SPAR-token pipeline (adjusted \(p=0.008\)). This is a
comparison of complete pipelines with different electrode
coverage and decoders, not an isolated quantization effect
(Fig.~\ref{S-fig:operating-points};
Section~\ref{S-app:n6}; Supplementary Data~1).

\subsection{Sensitivity to noise, electrode loss, calibration, and unseen recordings}
\label{sec:sensitivity}
\emph{Noise and electrode loss.}
Without retraining, we tested the decoders with simulated noise and electrode
loss after a clean calibration interval, and compared the increase in
error from clean to degraded inputs for tokens and continuous features.
In every condition, the pointwise upper 95\% confidence limit for the
token-minus-continuous difference in these increases was below 1\,mm
(largest 0.93\,mm, SPAR at 32\% noise), and three conditions showed smaller increases with
tokens (Holm-adjusted \(p\leq0.031\)). Nevertheless, severe
degradation affected SPAR more strongly than the learned
representation, with or without quantization
(Fig.~\ref{S-fig:noise}; Supplementary Data~8;
Section~\ref{S-app:corruption}).

\emph{Calibration.}
Replacing per-window calibration with calibration at
recording onset, without refitting, increased error by approximately
0.9\,mm. The pipelines had been fitted under per-window calibration,
so this tested a mismatch between training and evaluation.
After refitting standardization, codebooks, and decoders
under onset calibration, errors reached
10.3\,mm with tokens and 10.4\,mm with continuous SPAR,
both below the per-window results. In these short recordings,
calibration changes affected mean error more than
quantization did (Section~\ref{S-app:replays}).

\emph{Unseen recordings.}
Without retraining, SPAR decoders reached approximately 18\,mm on unseen
1.5\,Hz digit recordings, compared with 10\,mm on the
0.5\,Hz test windows, while the token-minus-continuous
difference remained small (\(+0.18\)\,mm). On seven additional
unseen movements, tokens had 0.2--0.7\,mm higher error
than continuous SPAR. The speed comparison changes both movement pace and whether the
recording contributed training windows, so neither effect can be
isolated (Section~\ref{S-app:replays}).

\section{Discussion}

Stored EMG tokens reduced the payload about 41-fold relative to
raw EMG and 36-fold relative to continuous SPAR features, with
small changes in hand-pose error. Averaged across decoders,
these changes were 0.01\,mm for SPAR and 0.45\,mm for the learned
representation (Fig.~\ref{fig:dpose-main}B--C). Both pooled
increases were below the 1\,mm bound. On emg2pose,
mean differences between learned tokens and continuous features
stayed below 0.05\,mm in 176 test participants, including 20
unseen during training (Fig.~\ref{S-fig:emg2pose-cohorts}B;
Supplementary Data~4). Across the evaluated pipelines,
representation choice, electrode coverage, calibration, and
unseen recordings had larger effects on accuracy than
tokenization (Figs.~\ref{fig:dpose-main}A
and~\ref{S-fig:operating-points}; Section~\ref{S-app:replays}).

Separating EMG encoding from pose decoding allows the decoder
to be updated while the wearable encoder and codebook stay fixed
(Fig.~\ref{fig:concept}B). Transmitting a decoded pose can use less
bandwidth, but places the task-specific decoder on the wearable.
Earlier EMG vector quantization focused on signal
reconstruction \citep{gronforsVectorQuantizationMethod2006}, and
BioCodec used discrete biosignal representations for downstream
tasks \citep{avramidis2025biocodec}. VQ-MyoPose learns tokens from
joint-angle trajectories and predicts them from EMG
\citep{lovecchio2025vqmyopose}. Here, the token representation is
fitted without pose labels, and the paired comparisons measure
the accuracy cost of tokenization alongside its payload and
computational costs.

With clean inputs, the SPAR-token pipeline had lower mean error than
the learned-token pipeline and met the 1\,mm non-inferiority criterion
relative to the adapted raw-waveform CNN (Fig.~\ref{fig:dpose-main}A). Under severe signal degradation
(strong noise or silenced electrode arrays), however, the learned
representation performed better
(Fig.~\ref{S-fig:noise}). The two pipelines also differed in
the data used to fit the tokenizer, input processing, and decoder training,
so their relative performance cannot be attributed to spatial
features alone. Within SPAR, each electrode's own band
power alone gave errors similar to the learned tokens, and replacing
these features with the full SPAR patch features reduced error by 2.5\,mm
(Section~\ref{S-app:ablation}).

Decoder complexity brought no consistent accuracy advantage.
With learned-token inputs, the compact spatial CNN + TCN had
lower mean error and fewer parameters than the TCN and Transformer
decoders. Qwen3's pretrained weights gave lower error than random
weights, but removing the backbone gave lower error with less
computation. TabPFN
reduced error by a further 0.37\,mm relative to the proposed
SPAR-token decoder, at a cost of over two minutes per pose when
every prediction reprocesses the training context (Fig.~\ref{fig:dpose-main}A;
Supplementary Data~2 and~3). The compact decoder offered
a more practical balance of accuracy and inference cost for
continuous pose estimation.

The retrieval analysis showed that tokens also retained useful
pose information without a trained decoder. Nearest-neighbour
retrieval error changed little when continuous features were replaced
by codebook vectors on held-out repetitions, whereas
matching token IDs by Hamming distance, which ignores distances
between codebook vectors, produced larger errors
(Fig.~\ref{fig:token-pose}D). Opening and closing were
also distinguishable at matched hand closure
(Section~\ref{S-app:manifold}). The trajectories in
Fig.~\ref{fig:token-pose}A--C visualize how tokens evolve over
repetitions, using token histories, movement-specific projections, and
phase-aligned repetitions.

Calibration and unseen recordings affected token and
continuous-feature decoders similarly (Section~\ref{S-app:replays}).
Changing calibration increased error, while refitting
standardization, codebooks, and decoders under the new calibration
improved both representations. Matching the calibration used in
training therefore mattered more than whether the decoder received
continuous features or tokens. On unseen recordings, absolute
errors rose substantially while the difference between tokens and
continuous features stayed small.

Under the assumed Bluetooth throughput, on-device computation limited
the channel count more than transmission did. The optimized SPAR implementation processed
63 channels within 20\,ms, while the learned tokenizer exceeded
that interval for one channel (Fig.~\ref{fig:stm32n6}). The
63-channel SPAR-token pipeline also had lower error than the
42-channel raw-EMG pipeline that fitted within the same throughput,
although electrode coverage and decoding differed
(Fig.~\ref{S-fig:operating-points}). The hardware benchmark timed
token generation. Acquisition, radio transmission, control,
and power consumption were outside its scope
(Section~\ref{S-app:n6}). Compact models such as TinyMyo
\citep{fasulo2026tinymyo} and embedded graphics processors
\citep{pfenningAnalysisEmbeddedGPU2024} offer an alternative
through local inference. Token transmission instead keeps task-specific
processing at the receiver, where the decoder can be changed
without changing the wearable representation.

The main HD-EMG results were obtained using personalized,
within-session decoders for nine participants. Learned-tokenizer pretraining excluded the
validation and test windows but included other windows from the same
participants and sessions (Section~\ref{S-app:data}), and SPAR
was evaluated within sessions with calibration from short
recordings. Future work should test whether a fixed codebook stays valid across
sessions, electrode re-donning, and clinical populations, and how
token transmission performs in closed-loop control on a complete,
power-limited wearable.

\section{Conclusion}
EMG tokenization provides a compact interface between wearable sensing and three-dimensional hand-pose decoding. Using two representations fitted without pose labels, we showed that substantial payload reductions are possible with small changes in decoding accuracy. In personalized, within-session experiments, stored token arrays reduced payloads approximately 41-fold relative to raw EMG, while mean error increases remained below 1\,mm when averaged across decoders. Learned tokens also performed similarly to continuous features on emg2pose, including participants unseen during training. Across the evaluated pipelines, representation choice, electrode coverage, calibration, and unseen recordings affected accuracy more than quantization.

The optimized SPAR implementation generated tokens for 63 channels within a 20\,ms update interval on a Cortex-M55 microcontroller. Together, these findings support tokenization as a means of reducing communication demands while retaining information useful for movement decoding. By separating wearable encoding from task-specific processing, this approach allows hand-pose decoders to be updated or replaced at the receiver while the wearable encoder and transmitted representation remain fixed.

\section*{Acknowledgment}
Funded by the European Union. Views and opinions expressed are
however those of the author(s) only and do not necessarily reflect
those of the European Union or the European Research Council
Executive Agency. Neither the European Union nor the granting
authority can be held responsible for them.

The authors gratefully acknowledge the scientific support and high-performance computing resources provided by the Erlangen National High Performance Computing Center (NHR@FAU) of the Friedrich-Alexander-Universität Erlangen-Nürnberg (FAU). The hardware is partially funded by the German Research Foundation (DFG).

\FloatBarrier
\printbibliography

\end{document}


\renewcommand{\thesection}{S-\Roman{section}}
\renewcommand{\thefigure}{S\arabic{figure}}
\renewcommand{\thetable}{S\arabic{table}}
\renewcommand{\theequation}{S\arabic{equation}}
\title{Supplementary Material for:\\EMG Tokenization Preserves Movement Information at a Fraction of the Bitrate}
\author{
  Raul C. S\^{i}mpetru \orcidlink{0000-0003-0455-0168},
  Amin Olamazadeh \orcidlink{0009-0002-6830-5559},
  Devon Rohlf \orcidlink{0000-0003-1601-0046},
  Vlad Cnejevici \orcidlink{0000-0003-2291-5941},
  and Alessandro Del Vecchio \orcidlink{0000-0002-7118-5554}
   \thanks{This work was supported in part by the European Research
Council (ERC) through the GRASPAGAIN project under Grant 101118089
and in part by the German Federal Ministry of Research, Technology
and Space (BMFTR) through the BIONIK project under Grant 16SV9639,
the EMGRoll project under Grant 16SV9306, and the NOE-EMY project
under Grant 16SV9246, all to A. Del Vecchio. (Corresponding author: Alessandro Del Vecchio)}
\thanks{The authors are with the Department of Artificial Intelligence in Biomedical Engineering, Friedrich-Alexander-Universität Erlangen-Nürnberg, 91052 Erlangen, Germany}
}
\maketitle
\makeatletter
\def\ps@headings{%
  \def\@oddhead{\hbox{}\@IEEEheaderstyle\leftmark\hfil\thepage/\pageref{LastPage}}\relax
  \def\@evenhead{\@IEEEheaderstyle\thepage/\pageref{LastPage}\hfil\leftmark\hbox{}}\relax
  \let\@oddfoot\@empty
  \let\@evenfoot\@empty
}
\def\ps@IEEEtitlepagestyle{%
  \def\@oddhead{\hbox{}\@IEEEheaderstyle\leftmark\hfil\thepage/\pageref{LastPage}}\relax
  \def\@evenhead{\@IEEEheaderstyle\thepage/\pageref{LastPage}\hfil\leftmark\hbox{}}\relax
  \let\@oddfoot\@empty
  \let\@evenfoot\@empty
}
\makeatother
\pagestyle{headings}
\thispagestyle{IEEEtitlepagestyle}
\raggedbottom
\clubpenalty=10000
\widowpenalty=10000
\displaywidowpenalty=10000

\section{Communication and embedded implementation}\label{app:contract}
\subsection{Payload and memory accounting}\label{app:bandwidth}
The payload rate depends on the channel count, update rate, and bits per
value. We generate tokens at 50\,Hz and represent each index with
10 bits when packed or 16 bits in the stored arrays, whereas raw
electromyography (EMG) consists of 16-bit samples at 2,048\,Hz. The assumed 1.4\,Mbit/s application throughput \citep{bluetoothsigLEprimer}
accommodates 42 raw-EMG channels or 2,800 packed-token channels before
application headers. Lower-layer Bluetooth
overhead is already included in this throughput assumption, and our 9-byte header per update adds a further 450\,B/s.

\begin{figure}[!t]
\centering
\includegraphics[width=\linewidth]{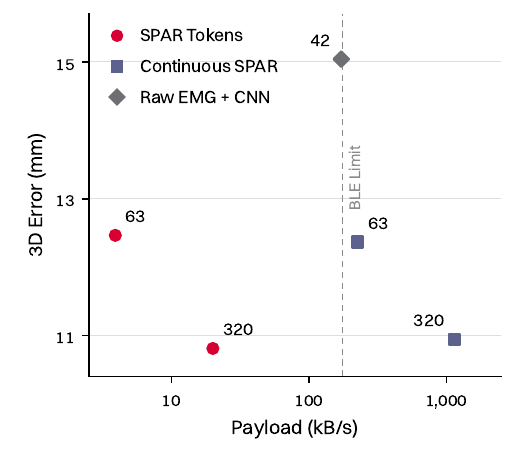}
\caption{\textbf{Electrode coverage, error, and payload.}
Points show mean three-dimensional (3D) pose error across nine participants and three decoder initializations, with labels indicating electrode counts. Payload excludes headers and uses ideal 10-bit packing for tokens, 32-bit floating-point values for the continuous spectral--spatial representation (SPAR), and 16-bit integers for raw electromyography (EMG). The dashed line marks the assumed 175\,kB/s Bluetooth LE (BLE) application throughput (BLE Limit). The 63-electrode SPAR and 42-electrode raw-EMG convolutional neural network (CNN) pipelines were refitted offline with different decoders. Supplementary Data~1 provides the full results.}

\label{fig:operating-points}
\end{figure}

Codebook storage is accounted for separately because it does not
contribute to the recurring payload. A 32-bit floating-point
(FP32) codebook for the spectral--spatial representation (SPAR)
occupies
72\,KiB for 1,024 entries or 4.5\,KiB for 64 entries, while the
learned codebook occupies 1\,MiB. Figure~\ref{fig:operating-points}
relates payload to offline decoding error at the tested electrode
counts, and Supplementary Data~1 lists the evaluated configurations,
including stored and packed payloads, token-generation time, and
decoder tensor memory. The units kB and MB are decimal, and KiB
and MiB are binary. The tensor
memory totals cover the listed model components rather than peak
memory during execution.

The emg2pose dataset contains 16 channels sampled at 2\,kHz and stored
as FP32 values, compared with the 320 channels of 16-bit EMG at
2,048\,Hz in the high-density EMG (HD-EMG) dataset of the main text.
On emg2pose, the learned-token stream requires 1.45\,kB/s including
per-update headers, compared with approximately 128\,kB/s for raw FP32
EMG and 820\,kB/s for FP32 continuous features
(Fig.~\ref{fig:emg2pose-comparison}B). The calibration statistics and
electrode descriptors used by the emg2pose decoder are sent separately,
in a setup message of about 2.3\,kB as text in our replay, and are not
included in these rates. Only its 32 calibration values, a centre and a
scale per channel, change between calibrations, which the per-window
evaluation protocol repeats for every 3\,s window.

\subsection{Timing measurement scope}\label{app:timing}
\emph{HD-EMG decoders.}
Except for TabPFN, decoder timing starts with inputs already in graphics processing unit (GPU) memory and
measures FP32 execution on an RTX 4090 graphics card (NVIDIA, Santa Clara, CA, USA) with automatic mixed
precision and TensorFloat-32 (TF32) disabled. Each call recomputes the input
context and predicts a complete hand pose. After warmup, we time
48 synchronized calls for each trained decoder and four
complete-pose calls for TabPFN, whose timing also includes central processing unit (CPU)
codebook lookup, feature summaries, principal component analysis
(PCA), transfers, and prediction
of every pose coordinate.

Fitting and loading are excluded from these timings. The reported
standard deviations describe variation between calls, while
parameter counts omit buffers, PCA, and retained training context
(Supplementary Data~3).

\emph{emg2pose end-to-end timing.}
This test replays recorded EMG through every processing step
needed to produce a pose prediction. We use one
prespecified 16-channel recording sampled at 2\,kHz and, after a
warmup run, time 100 updates covering preprocessing, transfers,
feature or token generation, packing and unpacking, codebook
lookup, and pose prediction. Live acquisition, network transmission,
and actuator time are excluded. For learned tokens, the median
time from raw EMG to pose was 4.4\,ms on a V100 GPU (NVIDIA) and 13.7\,ms
on four CPU threads of the same compute node (Xeon Gold 6134, Intel, Santa Clara, CA, USA, according to the cluster's hardware inventory), both below the
20\,ms update interval (Fig.~\ref{fig:emg2pose-comparison}C). TF32
was disabled during execution.

Fig.~\ref{fig:emg2pose-comparison}C omits BioCodec, whose implementation
re-encoded its full signal history at every update rather than
streaming. Learned-token
identifiers (IDs) were identical across repeated runs on each device, and 44 of
1,671,168 IDs (0.003\%) differed between CPU and GPU.

\subsection{Cortex-M55 and reduced-electrode validation}\label{app:n6}
\emph{Microcontroller timing.}
For Fig.~\ref{M-fig:stm32n6}, we measured token generation on
a NUCLEO-N657X0-Q board (STMicroelectronics, Geneva, Switzerland) running at 600\,MHz with FP32 arithmetic, caches
enabled, and no neural accelerator. Each configuration processes
deterministic synthetic input at 2\,kHz for 16 timed updates after
warmup. The benchmark excludes acquisition, radio transmission,
and pose decoding.

A configuration meets the timing budget only if every update
finishes within 20\,ms and its token checksum matches that of the
same C code running on the host computer. The original SPAR
implementation meets both conditions at 23 channels and the
optimized version at 63: the slowest 63-channel update takes
19.95\,ms, whereas 64 channels reach 20.28\,ms. The optimized
implementation uses the CMSIS fast real Fourier transform and a
square-root calculation for the hypotenuse in the Jacobi
eigensolver. Spectral feature extraction accounts for most of
the processing time, followed by codebook search
(Fig.~\ref{M-fig:stm32n6}B). The optimized learned tokenizer
requires 95.7\,ms for a single channel.

To vary workload size, we use the first $N$ electrode sites and
repeat the 320-site layout when $N$ exceeds 320. A patch member outside
the first $N$ sites is replaced by the centre electrode, and its coupling
is masked, so small workloads still form four-member patches. Band weights are
recalculated for each workload, but participant 1's 320-channel
standardization and codebook stay fixed. These workloads measure
processing cost. The reduced-electrode models used to assess
accuracy are fitted separately, as described below.

\emph{Numerical agreement.}
Alongside timing, we checked whether optimization preserved the
reference tokens and their decoded poses. On the nine participants'
320-channel test windows, optimized SPAR running on the host
reproduces 99.92\% of reference token IDs with negligible change
in mean pose error. Board and host outputs agree in the tested
configurations around the 63-channel operating point, and the full accuracy evaluation was run on the host.

\emph{Decoding with 63 electrodes.}
To measure the accuracy cost of reducing electrode coverage, we
fitted a separate offline model using the first forearm array
without its final corner electrode. Missing neighbours are
replaced by the electrode on the opposite side, and
standardization, codebooks, and decoders are refitted on training
data with absent decoder channels masked. The resulting errors
are 12.465\,mm with tokens and 12.368\,mm with continuous SPAR,
a paired difference of $0.097$\,mm (95\% confidence interval
$[-0.081,0.284]$\,mm; Holm-adjusted $p=0.26$). Both representations
have higher error than their 320-channel versions ($p=0.012$ each
in the same three-comparison family). Using the refitted
standardization parameters and codebooks, the optimized implementation running on the host computer reproduces 99.96\% of
reference token IDs.

\emph{Raw EMG within the same throughput.}
The corresponding raw-EMG control refits the adapted convolutional neural network (CNN) on the
first 42 electrodes, with its other inputs set to zero. Its mean
error exceeds that of the 63-electrode SPAR-token pipeline by
2.57\,mm (95\% confidence interval $[0.95,4.93]$\,mm; Holm-adjusted
$p=0.008$). Holm correction covers two comparisons: the 42-channel raw-EMG CNN versus the 63-channel SPAR-token pipeline, and the
42-channel versus the 320-channel raw-EMG CNN. Because electrode coverage
and decoder differ, these are comparisons of complete pipelines.

\emph{Reduced codebook size.}
We also fitted the 64-code SPAR configuration separately, and its
corresponding continuous-feature decoder achieved a mean error of
10.892\,mm. Tokens increase
error by 0.137\,mm (95\% confidence interval $[-0.132,0.391]$\,mm;
Holm-adjusted $p=0.258$ across its two comparisons with continuous
SPAR and learned tokens).

\section{Implementation and training}\label{app:training}
\subsection{Data separation}\label{app:data}
HD-EMG acquisition follows the main Methods and the dataset paper
\citep{simpetruLearningHandModel2024}. Learned-tokenizer pretraining
used recordings from all 13 participants, excluding the nine
evaluated participants' validation and test windows but retaining other
windows from those participants and sessions. SPAR standardization
and codebook fitting instead used only each participant's training
windows. In Section~\ref{app:exposure}, both tokenizers are fitted only on
the four participants outside the decoding cohort.

For the wrist-EMG emg2pose benchmark
\citep{salterEmg2poseLargeDiverse2024}, we formed deterministic
subsets of 3\,s windows from the official training, validation,
and test partitions. Within each partition, we selected recordings at
evenly spaced indices in metadata order and then evenly spaced windows
within each selected recording (at most 512 recordings and
8,192 windows for training, 256 and 2,048 for validation, and 512 and
4,096 for testing). Removing duplicate recording/start pairs left
7,849 training, 2,040 validation, and 4,095 test windows.
The test subset covers 176 participants, including 20 who were
unseen during both tokenizer pretraining and supervised decoder
training. Tokenizer pretraining used only official-training
recordings, although most other test participants contributed
different recordings to that pool.

The separate session evaluation used held-out same-day recordings
from ten participants. These recordings were excluded from both
tokenizer and decoder training, whereas other recordings from those
participants, including a session with the same hand and movement stage,
remained in the decoder training data and were available for
pretraining.

\subsection{Learned tokenizer}\label{app:tokenizer}
\emph{Signal normalization.}
For each 3\,s decoding window, the first 500\,ms provide a baseline
and scale for each EMG channel. We subtract the channel's lower
median and divide by $\max(1.4826\,\mathrm{MAD}, 10^{-4})$, in each
dataset's stored amplitude units, where MAD is the lower median absolute
deviation
\citep{rousseeuwAlternativesMedianAbsolute1993}, and clip normalized
samples at $\pm12$ robust scale units.

HD-EMG is resampled from 2,048 to 2,000\,Hz, with different filters
for pretraining and decoding. Pretraining uses zero-phase polyphase
resampling, whereas decoding uses a causal Kaiser finite impulse
response (FIR) filter so that it requires no future samples.
We retain the causal filter's 20\,ms delay and do not shift pose
targets to compensate for it.

\emph{Encoder architecture.}
The encoder in Fig.~\ref{M-fig:learned-tokenizer} processes
each EMG channel separately. A temporal convolution with 32 filters,
a kernel of 320 samples, and a stride of 40 reads 160\,ms of signal
every 20\,ms, and a Gaussian error linear unit (GELU) activation and pointwise convolution then
produce a 256-component feature vector.

Four residual blocks extend the temporal context. Each contains
layer normalization, a depthwise temporal convolution with kernel
5, and pointwise convolutions that expand the feature dimension
to 1,024 and reduce it to 256, with a GELU activation between
them and no dropout. The depthwise convolution processes each
feature separately and uses only current and preceding inputs.
After the residual blocks, a final layer normalization precedes
the nearest-vector search in the $1{,}024\times256$ codebook. The encoder has 2.13 million parameters and a 480\,ms receptive field.

Because this context starts after the 500\,ms calibration interval,
the first token with a complete receptive field becomes available
980\,ms after window onset, excluding computation.

\emph{Pretraining data and optimization.}
Pretraining alternates between HD-EMG and the emg2pose training
subset. The encoder first learns spectral reconstruction for
50,000 steps without quantization. After codebook initialization,
training continues for 250,000 quantized steps (seed 260905).
Both stages use AdamW with a constant learning rate of
$3\times10^{-4}$, $\beta=(0.9,0.999)$, weight decay 0.01, and
gradient-norm clipping at 10. We use 2\,s crops in batches of two,
reduced to one for sources with 64 or more channels, with FP32
arithmetic and TF32 matrix products allowed.

Input augmentation is applied after the reconstruction targets
are computed, so the targets stay unaugmented: each channel is multiplied by a log-normal gain
($\sigma=0.2$), and its polarity is flipped with probability 0.5.

\emph{Reconstruction objective and codebook updates.}
A two-layer reconstruction network, comprising layer normalization,
256 hidden units, GELU, and 64 outputs, predicts the EMG spectrum
from the encoder output. To construct its target, we subtract the
signal mean, apply a symmetric Hann window, and take the first
64 nonzero-frequency bins of $\log(1+|\mathrm{rFFT}(\cdot)|^2)$,
where rFFT is the Fourier transform for real-valued input.
The 160\,ms window gives 6.25\,Hz bin spacing and 64 bins up to
400\,Hz, with the zero-frequency bin omitted after mean removal. Predictions
are compared with these targets using smooth-$L_1$ loss.

During quantized training, a straight-through estimator passes
gradients through nearest-code selection \citep{oord2017vqvae}.
Encoder outputs are also kept close to their assigned codebook
vectors through the commitment loss
\begin{equation}
\mathcal{L}_{\mathrm{commit}}
=\frac{\beta}{D}\left\|z-\operatorname{sg}(e_q)\right\|_2^2,
\label{eq:commitment}
\end{equation}
where $z$ is the encoder output, $D=256$, $e_q$ is its assigned
codebook vector, and $\beta=0.25$. The stop-gradient operator
$\operatorname{sg}$ prevents this term from updating the codebook.
To discourage assignments from concentrating in too few entries,
we compute the entropy $H$ of the mean soft assignment using a
softmax of negative distances with temperature 0.05 and add
$0.1\,\max(0,\log 256-H)^2$. This term penalizes effective use of
fewer than 256 entries, a quarter of the codebook.

The codebook is initialized with Lloyd's $k$-means algorithm
(10 iterations) applied to sampled encoder outputs. During
quantized training, its vectors are updated by an exponential
moving average with decay 0.95, and entries whose moving-average count falls below 0.5 are reinitialized from current encoder
outputs. Reconstruction is scored only for tokens whose 160\,ms
input window ends after the 500\,ms calibration interval.
Inference additionally requires a complete 480\,ms context after
calibration.

\emph{Numerical precision and continuous controls.}
Each token--feature comparison uses matching computation settings.
Precomputed continuous features are stored in FP32 for emg2pose and the proposed HD-EMG decoder, whereas
the matched HD-EMG decoders and Qwen use features rounded to 16-bit
floating-point values with mixed-precision execution. TF32
convolutions are permitted during emg2pose feature precomputation, while
the FP32 runtime benchmarks disable TF32. In every paired control,
the continuous features come from the same encoder trained with
vector quantization (VQ), with
quantization skipped, rather than from a separately trained
encoder.

\subsection{SPAR feature extraction}\label{app:spar}
\emph{Neighbourhood construction.}
SPAR describes each electrode together with nearby electrodes in
the same grid. Electrode indices run row by row through three
$8\times8$ forearm grids and two $13\times5$ wrist grids, with
the final site of each wrist grid unrecorded. Each patch contains
four distinct electrodes: the centre, an auxiliary electrode,
and horizontal and vertical neighbours (Fig.~\ref{M-fig:spar-patch}).

The horizontal and vertical neighbours are normally to the right
and below the centre. At a grid boundary, a missing neighbour is
replaced by an electrode on the opposite side of the centre, using the left electrode instead
of the right or the upper electrode instead of the lower. The
auxiliary electrode is the unused left neighbour when available and otherwise
the unused site with the smallest sum of absolute row and column offsets from the centre,
with ties resolved first by row index and then by column index. Patches
never cross grid boundaries.

A substituted neighbour contributes to the spectral calculation,
but its retained coupling is set to zero after the matrix logarithm.
Across the 320 electrodes, this yields 270 horizontal and 286
vertical couplings, giving 5,256 valid features in a layout of
5,760 with 504 structural zeros.

\emph{Spectral features.}
SPAR uses the same causal resampling from 2,048 to 2,000\,Hz and the same per-window normalization as the learned tokenizer (Section~\ref{app:tokenizer}).
For each patch, the 160\,ms window is divided into seven overlapping
64\,ms segments starting 16\,ms apart. Each segment is multiplied
by a symmetric Hann window and Fourier transformed without zero
padding, after which the spectrum is grouped into six bands:
20--40, 40--60, 60--100, 100--160, 160--250, and 250--500\,Hz.
Each band includes its lower bound and excludes its upper bound.
At 2\,kHz sampling, the 128-point transform gives 15.625\,Hz bin
spacing, so these bands contain 1, 1, 3, 4, 5, and 16 bins,
respectively. The first two therefore contain only the 31.25 and
46.875\,Hz bins, respectively, but their cross-spectral estimates still average over all seven segments.

Within a band, frequency bins are weighted by their relative
contribution to power. We obtain these weights by averaging power
over segments, summing over recorded channels, and normalizing
within the band. The resulting cross-spectral matrix describes
the four electrodes and their spectral relationships. Before
computing its matrix logarithm by eigendecomposition, we enforce Hermitian symmetry
by averaging the matrix with its conjugate transpose, add $10^{-6}$ times its largest
diagonal entry to every diagonal element, then add $10^{-8}$ times the new largest
diagonal entry, and floor the eigenvalues at $10^{-12}$.

Each logarithmic matrix contributes three real features: the
centre's diagonal entry and the real parts of its couplings to
the horizontal and vertical neighbours. The six bands therefore
give 18 features per electrode. Because the logarithm is applied
to the matrix rather than to individual entries, the diagonal
feature is not simply log band power, and the features also depend on cross-spectral phase relationships.

\emph{Standardization and quantization.}
We standardize each of the 18 components using its mean and
population standard deviation across training electrodes and
updates, with the standard deviation floored at $10^{-4}$. Entries fixed at zero because of
missing neighbours are excluded from these statistics. These
structural zeros are restored before quantization and after
codebook lookup.

For each participant, $k$-means with 40 Lloyd iterations and seed
260911 fits 1,024 centroids to a random subset of training feature
vectors. The fitting budget is 128 vectors per centroid. Each
standardized vector is assigned to its nearest centroid using a distance calculation
that includes the structural zeros. We store codebooks as FP32
values and token indices as unsigned 16-bit integers (uint16).
The 64-entry codebook is fitted with the same procedure and seed, and the
pose-retrieval analyses use the same fitting procedure.

\subsection{HD-EMG pose decoders}\label{app:decoders}
Within each token-versus-continuous comparison, we match decoder
architecture, targets, initialization, and training procedure
(Table~\ref{tab:decoders}). Neural decoders are fitted with three
seeds whose errors are averaged within each participant, whereas
TabPFN keeps one pretrained network and varies its ensemble seeds.
The matched decoders, the spatial CNN with a temporal convolutional network (TCN), and Qwen predict
coordinates standardized with training-only means and population
standard deviations floored at $10^{-4}$. TabPFN and the adapted CNN use
coordinates in millimetres, and the CNN published in IEEE Transactions on Biomedical Engineering
\citep{simpetruLearningHandModel2024} and retrained here (TBME CNN) predicts training-only min--max-scaled
coordinates. Each 3\,s window supplies 100 updates, from 1.02 to 3.00\,s
after its start, and the final 50 are scored, so the first scored
prediction sees 51 updates and no learned representation with an
incomplete receptive field enters a decoder. Transformers and Qwen3 use
causal attention, TCNs pad causally with zeros, and TabPFN averages only
the available history.

\begin{table*}[!t]
\caption{\textbf{Pose decoders.} Architecture, training, and trainable parameters. Neural decoders use AdamW with weight decay 0.01 and gradient-norm clipping at 1 unless their row states otherwise. TCN, temporal convolutional network; CNN, convolutional neural network; GELU, Gaussian error linear unit; MLP, multilayer perceptron; SPAR, spectral--spatial representation; PCA, principal component analysis; TBME, IEEE Transactions on Biomedical Engineering; EMG, electromyography; LR, learning rate; SWA, stochastic weight averaging; MSE, mean squared error; FP32, 32-bit floating point; M, million.}\label{tab:decoders}
\footnotesize\setlength{\tabcolsep}{4pt}\renewcommand{\arraystretch}{1.1}\rowcolors{2}{black!7}{white}
\begin{tabularx}{\linewidth}{@{}>{\raggedright\arraybackslash}p{2.5cm}X>{\raggedright\arraybackslash}p{6.3cm}>{\raggedleft\arraybackslash}p{1.3cm}@{}}
\toprule
\rowcolor{white}Decoder & Architecture & Training and checkpoint & Parameters\\ \midrule
Matched TCN & Learned encoder features or codebook vectors; width 404; geometry-aware pooling (four slots, four heads); 7 causal blocks, kernel 2, dilations 1--64; temporal dropout 0.1; up to 100 updates (2\,s) of context & Smooth-$L_1$; 15,000 steps of two windows sampled with replacement, scoring the final second of each; 500-step warmup to LR $3\times10^{-4}$, cosine decay to $3\times10^{-6}$; seeds 260911--260913; earliest checkpoint with the lowest validation error; 16-bit mixed precision & 7.55\,M\\
Matched Transformer & As the matched TCN but width 468; 2 pre-normalization layers, 4 heads, feed-forward 936, causal mask & As the matched TCN & 7.45\,M\\
Matched waveform CNN & Adapted CNN architecture with dropout 0.1 & As the matched TCN, on 156.25\,ms waveform windows ending at the scored updates & 7.49\,M\\
Spatial CNN + TCN (proposed) & $1\times1$ convolution to 64; two spatial convolutions over five groups of 64 electrodes in acquisition order (32 filters, GELU, instance normalization; $5\times32$ with dilation $(1,2)$ and circular padding, then $5\times10$); 7 causal temporal blocks (kernel 2, dilations 1--64, MLP width 128, dropout 0.1); head $64{\to}128{\to}128{\to}60$; structural zeros reapplied for SPAR & Smooth-$L_1$; batches of two windows; LR $3\times10^{-4}$; up to 40 epochs, stopping after 6 without improvement; lowest validation error; FP32 & 704,700 (689,468 with SPAR)\\
Adapted waveform CNN \citep{simpetruProportionalSimultaneousRealTime2023,simpetruLearningHandModel2024} & 156.25\,ms raw EMG and a 20\,Hz low-passed copy, standardized per grid and window; dropout 0.20 & Coordinate $L_1$ in mm; AMSGrad, weight decay 0.32, no gradient clipping; 150 epochs of 10,000 endpoints, batch 64; input noise $\sigma=0.1$; OneCycle from $10^{-4}$ to $3.16\times10^{-3}$ until 75\,\%, then SWA with LR annealed to $10^{-4}$ in five epochs; lowest validation error among epoch checkpoints and averaged weights; 16-bit mixed precision & 7.49\,M\\
Retrained TBME CNN \citep{simpetruLearningHandModel2024} & Published architecture and preprocessing: 93.75\,ms windows, notch filter, raw and low-pass inputs, per-grid and per-window min--max normalization & Adapted recipe: $L_1$ on training-only min--max-scaled targets; AMSGrad; 50 epochs; batches of 64 accumulated to 512; OneCycle peaking at $10^{-2.5}$; original, 5\,dB-noise and magnitude-warped copies; earliest checkpoint with the lowest validation error; 16-bit mixed precision & 27.4\,M\\
Qwen3 variants & Adapter: electrode features to 64 dimensions, geometry-aware pooling (four slots, four heads), layer normalization, linear map to 1,024, layer normalization, scaling by 0.02; layer-normalized linear pose head; Qwen3-0.6B-Base frozen, as pretrained weights (token and continuous inputs), random weights (tokens), or removed (tokens, current update only); tokens are never read as text & Matched schedule (loss, steps, LR, seeds, and checkpoint as the matched TCN); only adapter and head trained; bfloat16 autocast with 32-bit trainable weights & 0.40\,M\\
TabPFN & SPAR codebook vectors or continuous features with trailing means over 100\,ms, 500\,ms, and 2\,s, reset at each window; structural zeros removed; PCA to 128 components fitted per participant and input on training data (seed 260921); one regressor per pose coordinate, eight ensemble members & No gradient-based fitting; labelled context of 3,200 training updates (the final second of 64 windows), whose features, like those of the queries, contain no lagged pose coordinates; ensemble seeds 260911--260913 & frozen\\
emg2pose Transformer & Identity or linear input map to width 256; geometry-aware pooling (eight slots); 2 layers, 4 heads, causal mask, dropout 0.1, sinusoidal position encoding, up to 100 updates (2\,s) of context & Coordinate MSE $+\,0.1\times$ velocity MSE; batches of 16 windows; LR $3\times10^{-4}$; up to 50 epochs, stopping after 8 without improvement; seeds 260826--260828; lowest participant-balanced validation error & 2.5\,M total\\
\bottomrule
\end{tabularx}
\end{table*}

\emph{Additional decoder inputs.}
Auxiliary inputs differ across decoders. The emg2pose decoder
receives calibration statistics and electrode descriptors, the
matched HD-EMG temporal decoders receive electrode-geometry
descriptors with constant calibration inputs, and Qwen receives
electrode-geometry descriptors without calibration inputs.
The proposed HD-EMG decoder and TabPFN use neither. Each electrode
descriptor has 14 components: centre coordinates (normalized grid
row, column, and grid index for HD-EMG; position on the unit circle
in acquisition order for the emg2pose band), an orientation
vector (unit length when known, zero otherwise), log spacing, two flags for known
coordinates and orientation, and a five-way measurement-type
indicator. Calibration inputs comprise each channel's log scale
relative to the across-channel mean log scale, that mean, and the channel's
calibration median divided by its scale, clipped at $\pm12$. A learned projection maps them to 32 values
that are concatenated with the pooled electrode features, and the
constant-calibration controls use centre 0 and scale 1. These inputs
are supplied only to the decoders, leaving the tokenizers unchanged.

\subsection{emg2pose decoder and quantizers}\label{app:codec}
The emg2pose comparisons use the data partitions and pretraining
data described in Section~\ref{app:data}. Results are reported in
Section~\ref{app:emg2pose}.

\begin{table*}[!t]
\caption{\textbf{Representations compared on emg2pose.} Bits per channel and update, assuming packed indices for tokens and 32-bit floating-point storage for continuous features, before headers. VQ, vector quantization; PCA, principal component analysis; FSQ, finite scalar quantization; GELU, Gaussian error linear unit; MSE, mean squared error; LR, learning rate; RMS, root mean square; EMG, electromyography; FP32, 32-bit floating point; M, million parameters.}\label{tab:reps}
\footnotesize\setlength{\tabcolsep}{4pt}\renewcommand{\arraystretch}{1.1}\rowcolors{2}{black!7}{white}
\begin{tabularx}{\linewidth}{@{}>{\raggedright\arraybackslash}p{2.5cm}>{\raggedright\arraybackslash}p{5.4cm}X>{\raggedleft\arraybackslash}p{1.3cm}@{}}
\toprule
\rowcolor{white}Representation & Configuration & Fitting & Bits\\ \midrule
Learned tokens & VQ-pretrained encoder; 1,024-entry codebook of 256-dimensional vectors & Section~\ref{app:tokenizer} & 10\\
Continuous features & Same encoder, quantization skipped & -- & 8,192\\
Quantized PCA & Reference encoder, four components; 4/4/4/4, 8/8/4/4, or 16/16/16/16 levels & Axes from sampled training vectors; uniform levels between the 0.1st and 99.9th training percentiles & 8, 10, 16\\
FSQ \citep{mentzer2024fsq} & Reference encoder; $256{\to}128{\to}4$ projection with GELU, bounded by tanh; 8/5/5/5 levels (1,000 combinations) & Straight-through rounding with a temporary $4{\to}128{\to}256$ reconstruction network, discarded afterwards; MSE, AdamW, LR $3\times10^{-4}$, weight decay 0.01, 5,000 steps of 1,024 vectors; validation reconstruction error selects the checkpoint & 10\\
Engineered $k$-means & 67 features of the mean-removed preceding 480\,ms: $\log(1+x)$ of the RMS, mean absolute value, mean absolute successive difference, and 64 Hann-windowed band powers & Training-only standardization; 256 or 1,024 centroids, 40 Lloyd iterations; no pretrained encoder & 8, 10\\
Engineered continuous & The same 67 features & Training-only standardization & 2,144\\
BioCodec \citep{avramidis2025biocodec} & First 1, 3, or 6 residual codebooks (256 entries of dimension 128) of one six-codebook codec with forward-only recurrence; EMG causally downsampled to 1\,kHz with its 10\,ms delay retained & Trained from scratch on the pretraining recordings: 40,000 FP32 steps of 32 single-channel 2\,s crops; Adam, LR $5\times10^{-4}$, $\beta=(0.5,0.9)$, 10,000-step warmup and cosine decay; original time-domain ($\times0.2$), frequency-domain, and commitment ($\times2$) losses; final checkpoint; 3.19\,M parameters & 8, 24, 48\\
Supervised continuous & Same encoder architecture, random initialization & Trained jointly with the decoder on pose labels, without spectral pretraining or quantization & 8,192\\
\bottomrule
\end{tabularx}
\end{table*}

All representations in Table~\ref{tab:reps} use the emg2pose Transformer
and training procedure specified in Table~\ref{tab:decoders}, with three
initializations. Learned
features and their codebook vectors enter the decoder directly. Learned
linear projections map the four values of PCA or finite scalar
quantization (FSQ), the 67 engineered
values, or the 128-dimensional sum of BioCodec codebook vectors to the
common input dimension. During pose inference, these representations
are used without reconstructing the original encoder output. PCA and FSQ use the fixed reference encoder
(seed 260905) and are fitted on each protocol's training partition
without pose labels. Only the paired learned-token comparison is
repeated across three pretrained encoders (seeds 260905--260907),
averaging their errors rather than their predictions. Coordinate and
velocity prediction errors are divided by the standard deviations of
the corresponding training targets before the loss is computed. BioCodec's training budget differs from the learned
tokenizer's 300,000 steps (Section~\ref{app:training}).

Pose targets are wrist-relative landmarks on the benchmark's
generic hand model, so their absolute errors are not directly
comparable with HD-EMG errors. We score 16 landmarks after
excluding the wrist and four fixed metacarpophalangeal (MCP)
anchors at the finger bases. Coordinates are normalized by the
generic hand model's mean wrist-to-MCP distance, and errors are
converted back to millimetres using that same distance. Mean
Euclidean landmark error is averaged across valid predictions
and evaluated landmarks within each participant, then across
participants with equal weight. Tokenization effects are paired
participant-level differences in token minus continuous-feature
error, with negative values favouring tokens.

\subsection{Statistical implementation}\label{app:statistics}
The statistical analyses use participants as the independent
observations, with one error difference per participant for each
comparison. Windows and training seeds contribute to these
participant summaries rather than increasing the sample size.
For HD-EMG, the tests enumerate all 512 sign-reversal patterns
across the nine participants.

\emph{Primary comparisons.}
Holm correction \citep{holm1979simple} is applied across four comparisons.
The first two test whether SPAR tokens and learned tokens increase
error by less than 1\,mm relative to their continuous features.
For each of these two, the participant's token-minus-feature difference
is first averaged with equal weight over the decoders using that
representation: the proposed decoder and TabPFN for SPAR, and
the proposed decoder, TCN, Transformer, and Qwen3 for the learned
representation. The third comparison is SPAR versus learned tokens
using the proposed decoder, and the fourth tests whether SPAR
tokens with that decoder exceed the mean error of the adapted
waveform CNN by less than 1\,mm.

Each bounded comparison is a non-inferiority test
\citep{piaggio2012consort}, using a one-sided permutation (sign-flip)
test of a mean difference of at least $\delta=1$\,mm against a
smaller mean. For these comparisons, we double each one-sided $p$ value before applying Holm correction, and a comparison
is significant when its adjusted $p$ value is below 0.05. All four comparisons have
Holm-adjusted $p=0.016$, the smallest value attainable with four
tests and nine participants. When the bound is reduced to
$\delta=0.5$\,mm, the learned-token comparison is no longer significant
(Holm-adjusted $p=0.71$), while the other two non-inferiority
comparisons and the SPAR-versus-learned comparison remain significant
(Supplementary Data~2).

\emph{Bound.}
The 1\,mm bound is a third of the 3\,mm mean localization error
reported for the reference kinematics
\citep{cakiciGeneralizedFrameworkStudy2022}. It is also comparable to
the change in a participant's error when the same decoder is
retrained with a different seed, which averages 0.2--0.3\,mm for
the proposed decoder and 0.6--1.0\,mm for the matched TCN and
Transformer.

\emph{Intervals.}
Confidence intervals are obtained by subtracting candidate mean
differences from the participant values and retaining those
candidates not rejected by the two-sided permutation (sign-flip)
test. This procedure assumes symmetry of the differences around
the tested mean. We use it for all nine-participant intervals in
the main text and supplement, including intervals around individual
models' mean errors in Fig.~\ref{M-fig:decoding}B and
Fig.~\ref{fig:noise}. For intervals around models' mean errors, the calculation uses
participant errors and assumes symmetry about their mean.

These intervals are pointwise and are not adjusted for multiple
comparisons. Unless a secondary test family is specified, comparisons
outside the primary family are reported as estimates with these intervals.

\emph{Sensitivity analysis.}
Supplementary Data~2 also reports a single-step studentized
max-$T$ correction \citep{romano2005exact} across all 76 HD-EMG
comparisons, which combine tokenization contrasts with baseline
comparisons. Each sign pattern was applied to all comparisons at
once, and the largest absolute studentized mean difference formed
the reference distribution.

For emg2pose, the 12-comparison family uses a Monte Carlo max-$T$
test with 10,000 sign patterns, including the observed pattern,
each applied consistently across comparisons. Pointwise confidence
intervals are formed from 10,000 bootstrap samples by resampling
participants with replacement. Intervals for the 10- and
20-participant cohorts are descriptive. Any additional test
families are specified alongside their results below.

Supplementary Data~2 contains all HD-EMG comparisons, including
those not plotted, and Supplementary Data~5 contains the 12
emg2pose comparisons.

\section{Quantization across participants and methods}\label{app:dpose}
\subsection{Tokenizers fitted on separate participants}\label{app:exposure}
\begin{figure*}[!t]
\centering
\begin{minipage}[t]{.485\linewidth}
\textbf{A}\quad Tokenizers fitted on separate participants\par
\includegraphics[width=\linewidth]{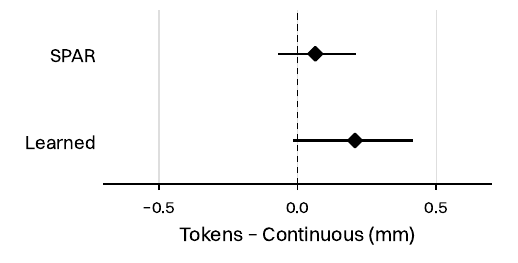}
\end{minipage}\hfill
\begin{minipage}[t]{.505\linewidth}
\textbf{B}\quad emg2pose test cohorts\par
\includegraphics[width=\linewidth]{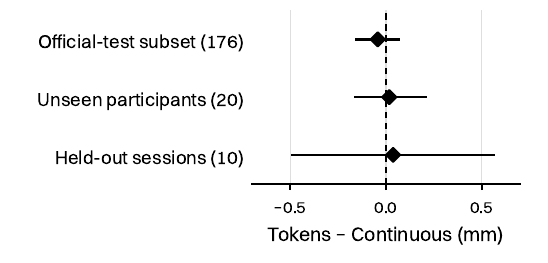}
\end{minipage}
\caption{\textbf{Quantization effects across participants.}
Points show mean token-minus-continuous error, with negative values favouring tokens.
\textbf{A:} Tokenizers were fitted on four separate participants, and personalized decoders were evaluated on nine participants. Bars show 95\% confidence intervals. Two-sided tests of a zero token-minus-continuous difference gave Holm-adjusted $p=0.32$ for the spectral--spatial representation (SPAR) and $p=0.12$ for the learned representation.
\textbf{B:} Errors are averaged across three pretrained encoders and three decoder initializations. Bars show pointwise 95\% participant-bootstrap intervals, with participant counts in parentheses. Unseen participants are part of the official test cohort, while held-out sessions were recorded on the same day. Supplementary Data~4 provides the values.}

\label{fig:quantization-transfer}\label{fig:emg2pose-cohorts}
\end{figure*}

To test whether small tokenization effects also occurred after
fitting tokenizers on different people, we fitted the learned
tokenizer and SPAR standardization and codebook exclusively on
four participants outside the nine-person decoding cohort, then
froze them. Personalized pose decoders were subsequently trained
for the nine evaluated participants using the main protocol.
Mean tokenization effects were small
(Fig.~\ref{fig:quantization-transfer}A). This analysis tests tokenizer
transfer across participants while retaining personalized pose
decoding.

\subsection{emg2pose comparisons}\label{app:emg2pose}
Across 176 test participants and three independently pretrained
encoders, learned tokens and continuous features achieved mean
errors of 28.00 and 28.04\,mm (Supplementary Data~6). Small paired mean differences
were also observed in the 20 participants unseen during tokenizer
and decoder training and in held-out same-day sessions from ten
participants (Fig.~\ref{fig:emg2pose-cohorts}B). The direction of
the effect varied across the three encoders. Alternative quantizers
that require encoder features use the reference encoder
(Section~\ref{app:codec}).

\begin{figure*}[!t]
\centering
\begin{minipage}[t]{.523\linewidth}
\textbf{A}\quad Decoding accuracy\par
\includegraphics[width=\linewidth]{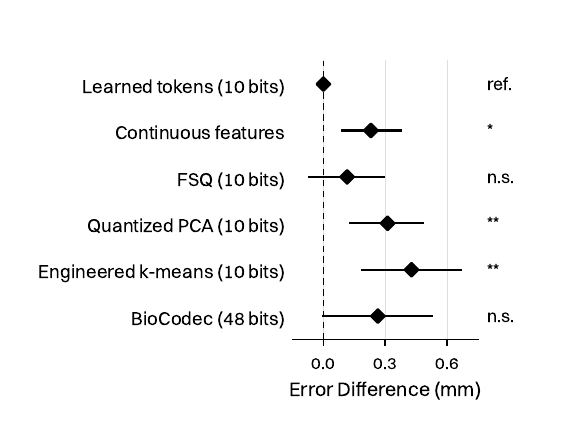}
\end{minipage}\hfill
\begin{minipage}[t]{.190\linewidth}
\textbf{B}\quad Transmission\par
\includegraphics[width=\linewidth]{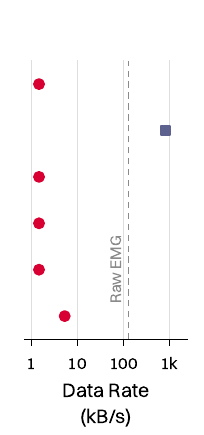}
\end{minipage}\hfill
\begin{minipage}[t]{.272\linewidth}
\textbf{C}\quad Computation\par
\includegraphics[width=\linewidth]{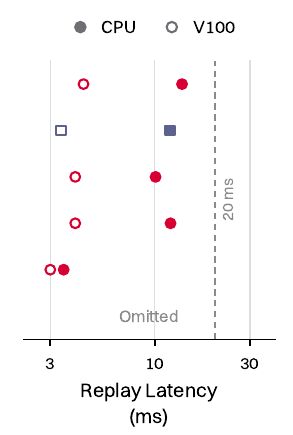}
\end{minipage}
\caption{\textbf{Accuracy, transmission rate, and processing time on emg2pose.}
Rows identify the same method across panels.
\textbf{A:} Error differences of finite scalar quantization (FSQ), quantized principal component analysis (PCA), and the other alternatives relative to learned tokens from the reference encoder. Positive values indicate higher error. Bars show pointwise 95\% participant-bootstrap intervals over 176 participants and three decoder initializations. Marks use max-$T$ correction across 12 comparisons: * $p<0.05$, ** $p<0.01$; n.s., not significant. This comparison uses one encoder, whereas Fig.~\ref{fig:emg2pose-cohorts}B averages three.
\textbf{B:} Data rates for 16 channels at 50\,Hz include 9-byte headers, with the dashed line marking raw electromyography (EMG) stored as 32-bit floating-point values at 128.45\,kB/s.
\textbf{C:} Median processing time from recorded EMG to pose output over 100 updates on a V100 graphics card (NVIDIA) and on four threads of a central processing unit (CPU), relative to the dashed 20\,ms interval. BioCodec timing is omitted (Section~\ref{app:timing}). Supplementary Data~5 provides all values.}

\label{fig:emg2pose-comparison}\label{fig:meta}
\end{figure*}

Figure~\ref{fig:emg2pose-comparison} compares these alternatives
in accuracy, transmission rate, and processing time. At 10 bits,
FSQ did not differ significantly from learned tokens, while
quantized PCA and engineered $k$-means had higher errors
(adjusted $p<0.01$). Separately trained BioCodec also showed no
significant error difference at its evaluated operating point,
but its six-codebook configuration uses 48 bits, outside the
matched 10-bit comparison.
FSQ is therefore a viable quantizer for features of a VQ-pretrained
encoder. The PCA and FSQ comparisons change the quantizer while keeping
the VQ-pretrained encoder fixed.

The held-out participant and same-day session tests apply to the
learned representation, while SPAR was evaluated only under the
within-session HD-EMG protocol. Supplementary Data~5 lists all
12 alternative-method comparisons with their joint max-$T$
correction, including settings not plotted, and Supplementary
Data~6 provides errors for all cohorts. HD-EMG comparisons and
decoder costs are reported in Supplementary Data~2 and~3.
Motion baselines are given in
Section~\ref{app:motion}, and payload and timing in
Section~\ref{app:contract}.

\subsection{SPAR feature ablation}\label{app:ablation}
To identify which SPAR features contribute to decoding, we refitted the nine-participant SPAR pipeline with six values per
electrode instead of 18. The first descriptor is the logarithm of
each electrode's own power in the six SPAR bands, computed with the
same frequency weights derived across all recorded channels but without
a four-electrode cross-spectral matrix or the matrix logarithm. The second keeps the six matrix-logarithm
diagonal entries and drops the couplings. Because each diagonal
entry depends on the whole patch matrix, it still reflects
neighbouring electrodes. Windows, standardization, 1,024-code
codebook fitting, decoder, seeds, and training are otherwise
unchanged (Section~\ref{app:spar}).

With tokens, mean errors were 13.31\,mm for log band power,
11.12\,mm for the matrix-logarithm diagonal, and 10.81\,mm for full SPAR. With continuous features, they were 13.06, 11.18, and
10.95\,mm. Relative to full SPAR tokens, log band power increased
error by 2.50\,mm (95\% confidence interval $[1.60, 3.36]$) and the diagonal by
0.31\,mm ($[0.15, 0.50]$; Holm-adjusted $p=0.008$ each). Most of the improvement over each electrode's own log band power
was retained with the matrix-logarithm diagonal alone, and adding the explicit couplings gave a smaller further improvement.
Relative to continuous features, tokenization changed mean decoding error by $+0.25$\,mm ($[0.02, 0.50]$) for log band power,
$-0.06$\,mm ($[-0.25, 0.12]$) for the diagonal, and $-0.13$\,mm
($[-0.32, 0.05]$) for full SPAR (Supplementary Data~9).

\section{Pose retrieval and movement direction}\label{app:manifold}
\subsection{Pose retrieval}\label{app:pose-retrieval}
For the analyses in this section, each recording is tiled with 3\,s
windows at a 2\,s stride. Each window is normalized from its own first
500\,ms and contributes 100 updates, which gives a continuous 50\,Hz
stream without duplicate updates.
Pose retrieval tests whether similar EMG representations correspond
to similar hand poses without training a pose decoder. We identify
movement cycles from peaks in low-pass-filtered measured mean
fingertip-to-wrist distance, using poses to define cycle membership and phase but not
to normalize features or fit codebooks.

Queries are updates whose 160\,ms analysis window lies inside a
test window (1,755--1,875 per participant), not only the final second
scored for the pose decoders. Candidates come from repetition cycles that do not overlap any
training window used to fit SPAR standardization and codebooks
(56--57 of about 150 cycles per participant). For each query update, we find
the closest representation in every eligible cycle across all eight
movements, excluding the query's own cycle and the two adjacent
cycles of the same movement. From these cycle-specific matches, we retain the five
closest to the query and average the measured
poses at those matches to estimate the query pose. Matching uses
squared Euclidean distance between full representations for
continuous SPAR and tokens restored to codebook vectors. Direct
matching of token IDs instead uses Hamming distance, defined as the
fraction of electrodes with different IDs, while the pose-space
reference selects matches by Euclidean distance between measured
pose coordinates.

To weight different parts of a movement equally, errors are first
averaged within ten phase bins per cycle and then across cycles,
movements, and participants. Continuous SPAR gives 7.92\,mm and
pose-space matching 4.38\,mm. The three comparisons in Fig.~\ref{M-fig:token-pose}D use exact permutation (sign-flip)
tests and intervals, with Holm correction across the tests:
$+0.09$\,mm with 1,024 codes (95\% confidence interval $[0.004, 0.18]$),
$+0.22$\,mm with 64 codes ($[0.08, 0.36]$), and $+0.92$\,mm for
Hamming matching of token IDs ($[0.54, 1.29]$; Holm-adjusted
$p=0.043$, 0.016, and 0.012; Supplementary Data~9).

\emph{All updates.}
Using every in-cycle update as a query, including those in the
training windows, and admitting candidate cycles that overlap training
windows, while still excluding the query's own and adjacent cycles, gives similar
token effects: $+0.07$, $+0.21$, and $+0.84$\,mm (Holm-adjusted
$p=0.012$ each).

The shifted control retains the token-based matches but repeatedly
shifts their associated poses at random within each cycle.
Its 11.3\,mm error is summarized as the mean of participant
medians, whereas the main retrieval results use the phase-balanced
means above. Under the same median-based summary, 1,024-code tokens
give 7.3\,mm. The shifted control and pose-space matching
are descriptive references, not performance limits.

\subsection{Movement-direction analysis}\label{app:direction}
The direction analysis asks whether the representation distinguishes
opening from closing at a similar degree of hand closure.
We define closure as the mean landmark distance from the movement's
open-hand template, rescaled between that movement's 1st and 99th
percentiles and clipped to $[0,1]$. Direction labels follow the sign
of the closure gradient after a centred five-update moving average
within each cycle, excluding updates with zero gradient.

Every in-cycle update of the eight movements serves as a query.
For each query, candidates are drawn from the same movement in
other nonadjacent cycles and must have normalized closure within
0.05 of the query. The closest candidate in the chosen representation
supplies the predicted direction. A majority-direction control
uses the same candidates but always predicts their more frequent
direction, choosing closing on ties. We average accuracy over the eight movements within
each participant.

Dequantized tokens achieve 70\% accuracy, compared with 55\% for
the majority-direction control, 65\% for matching measured poses,
and 64\% for Hamming matching of token IDs. Tokens exceed the
majority control and pose-based matching with Holm-adjusted
$p=0.012$ and $p=0.016$, respectively. The third comparison in
this family is token IDs versus pose-based matching. In a separate
comparison, continuous SPAR reaches 70.6\%, exceeding tokens by
0.6 percentage points ($p=0.004$).

\subsection{Visualization of token histories}\label{app:umap}
\emph{Input and embedding.}
The visualization in Fig.~\ref{M-fig:token-pose}A--C adds
temporal context to the representations used in the quantitative
analyses. It contains 14,861 updates from participant 1's eight
movement recordings, each with 320 tokens from the 1,024-entry
SPAR codebook. We replace each token with its 18-component
codebook vector, reapply structural zeros, and concatenate the
current vectors with those 6, 12, 18, and 24 updates earlier.
At 50\,Hz these lags span 480\,ms. History is taken from this
continuous stream and may cross window and cycle boundaries, each update
keeping its own window's calibration, without wrapping within a repetition.

After dropping constant columns, we fit one three-dimensional Uniform Manifold Approximation and
Projection (UMAP) \citep{mcinnesUMAPUniformManifold2018a}
across all movements, with no additional scaling or PCA. The
fit uses cosine distance, 50 neighbours, minimum distance 0.8,
spread 1.0, repulsion strength 0.05, 500 epochs, and random seed
42. Movement labels and measured poses enter only the subsequent
display construction, not the embedding fit itself.

\emph{Cycle phase and movement-specific projection.}
We construct the display by identifying repetition phase and then
choosing a projection plane for each movement. The measured mean
fingertip-to-wrist distance is detrended and filtered forwards and backwards at 50\,Hz
with a fourth-order 1\,Hz Butterworth filter. Consecutive peaks
with prominence of at least 0.2\,mm and an intervening trough
define a cycle. Within each recording, we retain durations between
the median divided by 1.8 and the median multiplied by 1.8.
Phase increases linearly from zero to one between the retained
cycle boundaries, near the open-hand pose.

For each movement, we average three-dimensional embedding
coordinates within 20 equal phase bins per cycle and then across
cycles with all bins populated. The first two singular vectors
of the centred mean trajectory define the plane onto which that
movement's points are projected.

The projected groups are positioned separately. We normalize the
original history vectors to unit length, average them by movement,
and normalize the resulting centroids. Two-dimensional metric
multidimensional scaling of their cosine distances initializes
group centres using seed 42 and eight initializations. These
centres are then rescaled and pushed apart until circles containing
each group's 98th-percentile radius no longer overlap, with a
gap of 0.12 times the mean radius. Because this placement is part
of the display construction, distances between groups are not
interpreted.

\emph{Colour and displayed points.}
Panel A colours points by movement label, while panel B colours
them by relative hand closure. For the latter, we calculate the
mean displacement of the 20 measured landmarks from a
movement-specific open-hand template defined by the first and
last 5\% of its cycles. Displacement is divided by the movement's
95th percentile and clipped at 100\%. All 14,861 updates are displayed, and axis limits span the 0.5th to 99.5th coordinate
percentiles with 5\% padding and contain every update.

\emph{Repetition curves.}
Panel C summarizes the phase-aligned repetitions as individual
curves and a mean trajectory. For the mean trajectory, each cycle contributes median coordinates in
20 equal phase bins, and the 19 complete cycles per movement
are averaged bin by bin. Individual repetition lines instead use
medians in ten phase bins and are omitted if a bin is empty or
an adjacent-bin jump, including the closing segment, exceeds
0.6 times the movement's mean coordinate range. This leaves
148 of 152 individual lines, while all 152 cycles contribute to
the mean trajectories.

We smooth the bin coordinates with circular Gaussian kernels over
offsets $-3$ to $+3$, using standard deviations of 0.5 bins for
mean trajectories and 0.8 bins for individual repetitions. Periodic
cubic splines then connect the smoothed coordinates, repeating
the first point at phase one and sampling 200 positions per curve.
Dots mark phase zero, and arrows follow increasing phase near
0.25 and 0.75. Periodic interpolation closes every curve by
construction.

\begin{figure*}[!t]
\centering
\includegraphics[width=\linewidth]{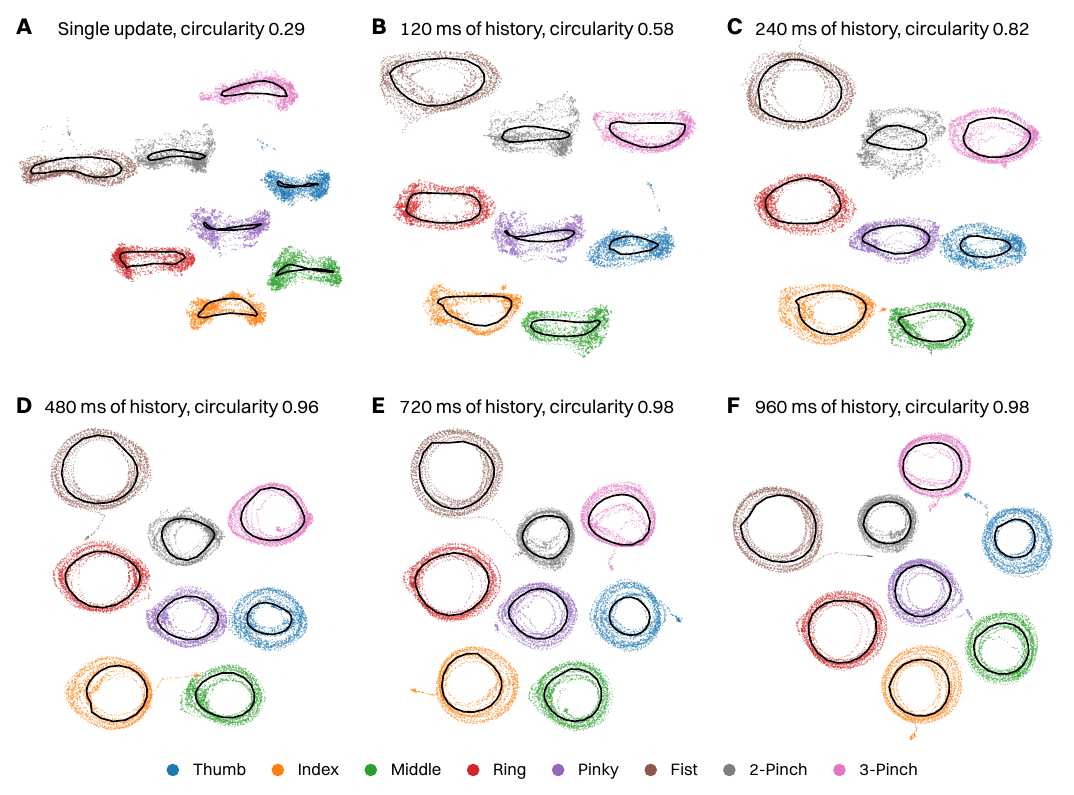}
\caption{\textbf{Effect of temporal context on the Uniform Manifold Approximation and Projection (UMAP) display.}
The same updates from participant 1 are embedded with increasing amounts of token history, using the UMAP settings, movement-specific projection planes, and group-placement procedure described in Section~\ref{app:umap}.
\textbf{A:} Current tokens alone.
\textbf{B--F:} Current tokens together with lagged updates every 120\,ms, extending back to 120, 240, 480, 720, or 960\,ms.
Colours identify the eight movements. Black lines connect the mean coordinates of 20 equal phase bins, averaged across complete repetitions, with the last bin connected to the first.
Panel titles give circularity $4\pi A/P^2$, averaged over movements, where $A$ and $P$ are the area and perimeter of these projected trajectories and a circle has circularity one. Circularity changes little beyond the 480\,ms used in Fig.~\ref{M-fig:token-pose}.
Each panel is fitted and arranged separately, so coordinates are not comparable across panels.}
\label{fig:token-history}
\end{figure*}

\emph{Temporal-context controls.}
Two controls examine how temporal context contributes to the
display. The history-length control repeats the embedding and
display procedure with 0, 120, 240, 480, 720, and 960\,ms of
history, adding lagged updates at 120\,ms intervals. Mean trajectory
circularity ($4\pi A/P^2$ for enclosed area $A$ and perimeter $P$, one
for a circle) rises from 0.29
for a single update to 0.96 at 480\,ms, with little further change
at 720 or 960\,ms (Fig.~\ref{fig:token-history}, panel titles).
Fig.~\ref{M-fig:token-pose} uses the 480\,ms history. To examine whether each history must stay paired
with its own current update, the shuffled-history control keeps the current update fixed but
reassigns the intact four-lag history blocks among updates within the
same repetition, then refits UMAP and recomputes each movement's projection plane with the same procedure, keeping the original cycle assignments and group anchors.
Mean trajectory circularity then falls from 0.96 to 0.03. The pose-retrieval and direction-classification
results use single updates in the original representations,
without this embedding or curve construction.

\section{Robustness and motion analyses}\label{app:noise}
\subsection{Sensitivity to noise and electrode loss}\label{app:corruption}
To compare sensitivity to degraded input, the learned, SPAR, and
adapted-CNN pipelines receive the same degraded recordings
without retraining, while the initial calibration interval stays
clean. Participant errors are averaged over three decoder initializations
and three independently generated degradation instances.

\begin{figure*}[!t]
\centering
\includegraphics[width=\linewidth]{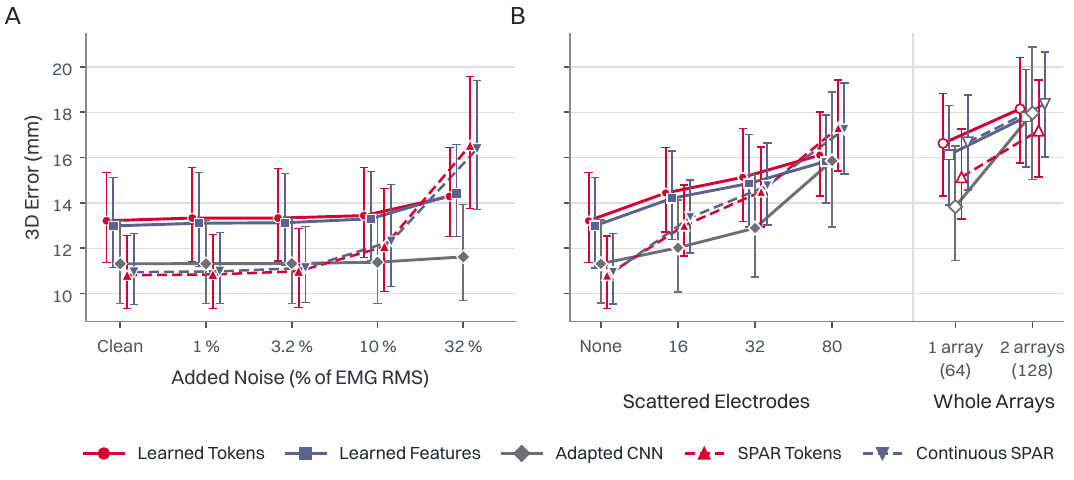}
\caption{\textbf{Decoding error under noise and electrode loss.}
Decoders for the learned representation, the spectral--spatial representation (SPAR), and the adapted convolutional neural network (CNN) are held fixed during evaluation, following the training procedures in Table~\ref{tab:decoders}. Points show mean three-dimensional (3D) error across nine participants after averaging three decoder initializations and three degradation instances, with 95\% confidence intervals.
\textbf{A:} The standard deviation of the added Gaussian noise is expressed relative to the root mean square (RMS) of each channel's electromyography (EMG).
\textbf{B:} Scattered electrodes or complete arrays are set to zero after calibration. Curves show absolute, unfiltered errors, whereas tests compare increases from each model's clean baseline. SPAR comparisons are reported in Supplementary Data~8.}

\label{fig:noise}
\end{figure*}

We add zero-mean Gaussian noise with standard deviations of 1, 3.2, 10, or 32\% of
each channel's EMG root mean square (RMS) after the calibration interval. Electrode loss is simulated by setting
16, 32, or 80 scattered channels to zero or by silencing one or
two complete 64-electrode arrays (Fig.~\ref{fig:noise}). None of
the models was trained with dropped channels. For each model,
we subtract clean-input error from degraded-input error and
compare the resulting increases within participants. The tests
therefore measure whether degradation increases error more with
tokens than with continuous features, while the figure shows
absolute errors.

At each noise level, we compare the increase in decoding error for
learned tokens with the corresponding increases for continuous learned
features and the adapted CNN.
These eight noise comparisons and the five electrode-loss comparisons are Holm-corrected in separate families.
At 32\% noise, the token decoder's error increases by 0.351\,mm
less than the continuous decoder's ($p=0.031$), although the
interval for their absolute error difference includes zero.
Under electrode loss, their error increases differ by at most
0.27\,mm,
with all upper 95\% confidence limits below 0.8\,mm.

SPAR uses separate correction families for four noise comparisons
and five electrode-loss comparisons. Its token errors increase by
1.40\,mm less than continuous-feature errors when one array is
silenced and by 1.03\,mm less when two arrays are silenced
($p=0.020$ for each). Supplementary Data~8 provides every SPAR
condition, interval, and test.

\subsection{Calibration and unseen recordings}\label{app:replays}
\emph{Calibration at recording onset.}
The main protocol estimates calibration statistics from the first
500\,ms of every input window. We instead use the first 500\,ms of each approximately 40\,s task
recording for calibration. We evaluate this change both without refitting and after refitting
the full SPAR pipeline under recording-onset calibration.

Without refitting, recording-onset calibration raises SPAR
error to 11.688\,mm with tokens and 11.736\,mm with continuous
features. Because
these models were fitted using per-window calibration, this
comparison introduces a mismatch between training and evaluation.

Refitting SPAR standardization, codebooks, and pose decoders
under recording-onset calibration gives errors of 10.271\,mm
with tokens and 10.448\,mm with continuous SPAR. Both improve
on the per-window results (Holm-adjusted $p=0.012$ and $p=0.016$,
respectively), and the token-minus-continuous difference is
$-0.177$\,mm (95\% confidence interval $[-0.336,-0.030]$\,mm; $p=0.031$
in the same three-test family). In both versions of this analysis,
each calibration applies to one task recording rather than an
entire session.

\emph{Unseen recordings.}
The original pipelines fitted under per-window calibration are also
evaluated on the faster 1.5\,Hz digit recordings, which were excluded from fitting. Each recording
is divided into overlapping 3\,s windows starting every 2\,s,
with per-window calibration and without retraining. Tokens give
18.31\,mm error and continuous SPAR gives 18.13\,mm, compared
with 9.98 and 10.14\,mm, respectively, on the main 0.5\,Hz digit
test windows. The token-minus-continuous difference is $+0.18$\,mm
(95\% confidence interval $[0.01,0.36]$\,mm; Holm-adjusted $p=0.070$). Holm correction covers the
token-minus-continuous contrasts in this evaluation and in the
recording-onset-calibration evaluation without refitting.

On seven further unseen circular, lateral, and vertical
movements, errors rose for both representations (16.3--47.6\,mm),
while tokens added 0.17--0.71\,mm relative to continuous SPAR.

\subsection{Motion and displayed trajectories}\label{app:motion}
\emph{Constant-prediction baseline and velocity.}
For emg2pose, the constant-prediction baseline holds each model's final prediction
before the scoring interval fixed throughout that interval.
It is not initialized with the measured hand pose. Learned tokens achieve
28.00\,mm on the sampled official-test windows, compared with 29.65\,mm for their
constant-prediction baseline, while continuous features achieve 28.04\,mm compared
with 29.63\,mm for their constant-prediction baseline (Supplementary Data~7).

Velocity is estimated from consecutive valid poses at 50\,Hz and
scored with a pooled $R^2$ against a constant mean-velocity
reference, weighting observations rather than participants.
Scores range from $-0.020$ to $+0.009$ across representations
(Supplementary Data~7): velocity errors are similar to those of
the constant reference for tokens and continuous features alike.
Predictions are not shifted in time before errors are calculated.

\emph{Displayed movement traces.}
Fig.~\ref{M-fig:decoding} illustrates predictions for
participant 6, whose SPAR-token error is just below the cohort
median. Predictions from additional input windows within the test block fill
the gap between the two scored one-second intervals to produce a continuous
four-second trace. The plotted distance depends on the movement:
thumb tip to pinky base for the thumb, fingertip to wrist for the
other digits, mean fingertip-to-wrist distance for the fist, and
thumb to fingertip for pinches, averaged over index and middle
for the 3-pinch.

For display, predictions are smoothed with a causal One Euro
filter \citep{casiez1FilterSimple2012} selected on learned-token
validation data. Its minimum cutoff is 1\,Hz, $\beta=0.1$, and
derivative cutoff is 1\,Hz at 50\,Hz, with the same settings
applied to SPAR traces. All numerical errors are calculated from
unfiltered predictions.

\section*{Data and code availability}
The HD-EMG recordings are available from the authors on reasonable
request \citep{simpetruLearningHandModel2024}, and emg2pose is
publicly available \citep{salterEmg2poseLargeDiverse2024}.
Supplementary Data~1--9 provide summary results and plot values.

Code is available at \url{https://github.com/NsquaredLab/MyoJi}.
\printbibliography